\documentclass[12pt,onecolumn,oneside,a4paper,peerreview]{IEEEtran}
\usepackage{cite}
\usepackage{graphicx}
\usepackage{amsmath}
\usepackage{times}
\usepackage{latexsym}
\usepackage{graphicx}
\usepackage{bm}
\usepackage{amssymb}
\usepackage[center]{caption2}
\usepackage{stfloats}
\usepackage{cases}
\usepackage{subfigure}
\usepackage{array}
\usepackage{setspace}
\usepackage{fancyhdr}
\usepackage{cite}

\newtheorem{theorem}{Theorem}
\newtheorem{lemma}{Lemma}

\def\proof{\noindent\hspace{2em}{\itshape Proof: }}
\def\endproof{\hspace*{\fill}~$\square$\par\endtrivlist\unskip}

\newcaptionstyle{mystyle2}{%
\captionlabel $.$ \, \doublespacing \captiontext \par}
\captionstyle{mystyle2}

\begin{document}
\title{Multi-antenna Wireless Legitimate Surveillance Systems: Design and Performance Analysis}
\author{Caijun Zhong, \emph{Senior Member}, \emph{IEEE}, Xin Jiang, Fengzhong Qu,  \emph{Member}, \emph{IEEE}, and  Zhaoyang Zhang, \emph{Member}, \emph{IEEE}
\thanks{C. Zhong, X. Jiang and Z. Zhang are with the Institute of Information and Communication Engineering, Zhejiang University, China (email: caijunzhong@zju.edu.cn). F. Qu is with the Ocean College, Zhejiang University, China.}}

\maketitle
\makeatletter
\newcommand{\rmnum}[1]{\romannumeral #1}
\newcommand{\Rmnum}[1]{\expandafter\@slowromancap\romannumeral #1@}
\makeatother

\begin{abstract}
To improve national security, government agencies have long been committed to enforcing powerful surveillance measures on suspicious individuals or communications. In this paper, we consider a wireless legitimate surveillance system, where a full-duplex multi-antenna legitimate monitor aims to eavesdrop on a dubious communication link between a suspicious pair via proactive jamming. Assuming that the legitimate monitor can successfully overhear the suspicious information only when its achievable data rate is no smaller than that of the suspicious receiver, the key objective is to maximize the eavesdropping non-outage probability by joint design of the jamming power, receive and transmit beamformers at the legitimate monitor. Depending on the number of receive/transmit antennas implemented, i.e., single-input single-output, single-input multiple-output, multiple-input single-output and multiple-input multiple-output (MIMO), four different scenarios are investigated. For each scenario, the optimal jamming power is derived in closed-form and efficient algorithms are obtained for the optimal transmit/receive beamforming vectors. Moreover, low-complexity suboptimal beamforming schemes are proposed for the MIMO case. Our analytical findings demonstrate that by exploiting multiple antennas at the legitimate monitor, the eavesdropping non-outage probability can be significantly improved compared to the single antenna case. In addition, the proposed suboptimal transmit zero-forcing scheme yields similar performance as the optimal scheme.
\end{abstract}

\section{Introduction} \label{section:1}

Wireless communications provide an efficient and convenient means for establishing connections between people. However, due to the open and broadcast nature of the wireless medium, wireless communications are particularly susceptible to security breaches, hence establishing reliable and safe connections is a challenging task. Responding to this, physical layer security, as a promising technique to enable secure communications, has attracted considerable attentions in recent years \cite{Z.Ding,N.Yang,J.Zhu1,J.Zhu2,Y.Zou2,Y.Zou3,F.Zhu,S.Gong,X.Jiang,X.Jiang1,X.Fang,F.Qahtani,Y.Huang,J.Zhu3,J.Zhu4}, and various sophisticated techniques such as artificial noise \cite{S.Goel,X.Zhou} and security-oriented beamforming \cite{A.Mukherjee,C.Jeong} have been proposed to enhance the secrecy performance.

In the physical layer security framework, the eavesdroppers are illegitimate adversaries, who intend to breach the confidentiality of a private conversation. On the other hand, wireless communications also facilitate the collaboration between the criminals or terrorists, thereby posing significant threats on national security. Therefore, to prevent crimes or terror attacks, there is a strong need for the government agencies to legitimately monitor any suspicious communication links to detect abnormal behaviors, such as communications containing sensitive word combinations, addressing information, or other factors with a frequency that deviates from the average.

For wireless communication surveillance, passive eavesdropping, where the legitimate monitor simply listens to the suspicious links, is a straightforward method. However, the legitimate monitor may be in general deployed far away from the suspicious transmitter to avoid getting exposed, as such the quality of the legitimate eavesdropping channel is a degraded version of the suspicious channel, making passive eavesdropping an inefficient approach. To circumvent this issue, a novel approach, namely proactive eavesdropping via cognitive jamming, was proposed in \cite{J.Xu,J.Xu1}, where the legitimate monitor, operating in a full-duplex manner, purposely transmits jamming signals to moderate the suspicious communication rate to improve the eavesdropping efficiency. Later in \cite{Y.Zeng}, the authors proposed three possible spoofing relay strategies to maximize the achievable eavesdropping rate.

In order to enable full-duplex operation, the legitimate monitor is equipped with two antennas, one for eavesdropping and the other for jamming. Also, an ideal assumption, namely, perfect self-interference cancellation, is adopted in \cite{J.Xu,J.Xu1}. However, residual self-interference is likely to exist due to practical constraints such as hardware impairment \cite{G.Zheng}. Under this realistic scenario, how to properly handle the self-interference becomes a critical issue to be tackled. In this paper, we propose to adopt multiple antennas at the legitimate monitor for performance enhancement. The motivation of using multiple antennas is two-fold, namely, enabling self-interference mitigation in the spatial domain and adjusting the effective jamming power observed at the suspicious receiver. For the considered multi-antenna wireless legitimate surveillance systems, we study the optimal joint design of jamming power and beamforming vectors. To reduce the complexity, intuitive suboptimal beamforming schemes are also proposed, and the achievable eavesdropping non-outage probability of the proposed schemes are examined.

The main contributions of this paper are summarized as follows:
\begin{itemize}
\item Depending on the number of receive/transmit antennas implemented at the legitimate monitor, i.e., single-input single-output (SISO), single-input multiple-output (SIMO), multiple-input single-output (MISO) and multiple-input multiple-output (MIMO), four different scenarios are studied. For each case, the optimal jamming power is derived in closed-form. In addition, employing the semidefinite relaxation (SDR) technique, the efficient algorithms are obtained for the optimal transmit/receive beamforming vectors.
\item Three low-complexity suboptimal beamforming schemes are proposed, namely, transmit zero-forcing (TZF)/ maximum ratio combing (MRC), maximum ratio transmission (MRT)/ receive zero-forcing (RZF), and MRT/ MRC. Closed-form expressions for the eavesdropping non-outage probability of TZF/MRC and MRT/RZF schemes are derived. In addition, simple and informative high SNR approximations of all suboptimal schemes are presented.
\item The findings of the paper suggest that, deploying multiple antennas is an effective means to enhance the system performance. Also, the optimal joint jamming power and beamforming scheme outperforms the proposed suboptimal schemes, the performance gap is rather insignificant compared with the TZF/MRC scheme, and gradually diminishes when the maximum jamming power becomes large. In addition, full diversity can be achieved by the MRC scheme, while the RZF attains a lower diversity since one degree of freedom is used for self-interference cancellation.
\end{itemize}

The remainder of the paper is organized as follows. Section \ref{section:2} describes in detail the system model and formulates the optimization problem. Section \ref{section:3} presents the optimal solutions. Suboptimal schemes are proposed in Section \ref{section:4}. Numerical results and discussions are given in Section \ref{section:5}. Finally, Section \ref{section:6} concludes the paper and summarizes the key outcomes.

{\it Notation}: We use bold upper case letters to denote matrices, bold lower case letters to denote vectors and lower case letters to denote scalars. $||\cdot||$, $(\cdot)^\dagger$, $(\cdot)^{-1}$ and $\text{tr} (\cdot)$ denote Euclidean norm, conjugate transpose operator, matrix inverse and the trace of a matrix, respectively. ${\tt E}\{x\}$ stands for the expectation of the random variable $x$ and ${\mathop{\rm Prob}\nolimits}(\cdot)$ denotes the probability. ${{\bf{I}}_k}$ is the identity matrix of size $k$. $O(\cdot)$ denotes the infinitesimal of the same order. $\Gamma(x)$ is the gamma function \cite[Eq. (8.31)]{Tables} and $\Gamma \left( {\alpha ,x} \right)$ is the upper incomplete gamma function \cite[Eq. (8.350.2)]{Tables}.

\section{System Model and Problem Formulation} \label{section:2}
\begin{figure}[htbp]
\centering
\includegraphics[width=4in]{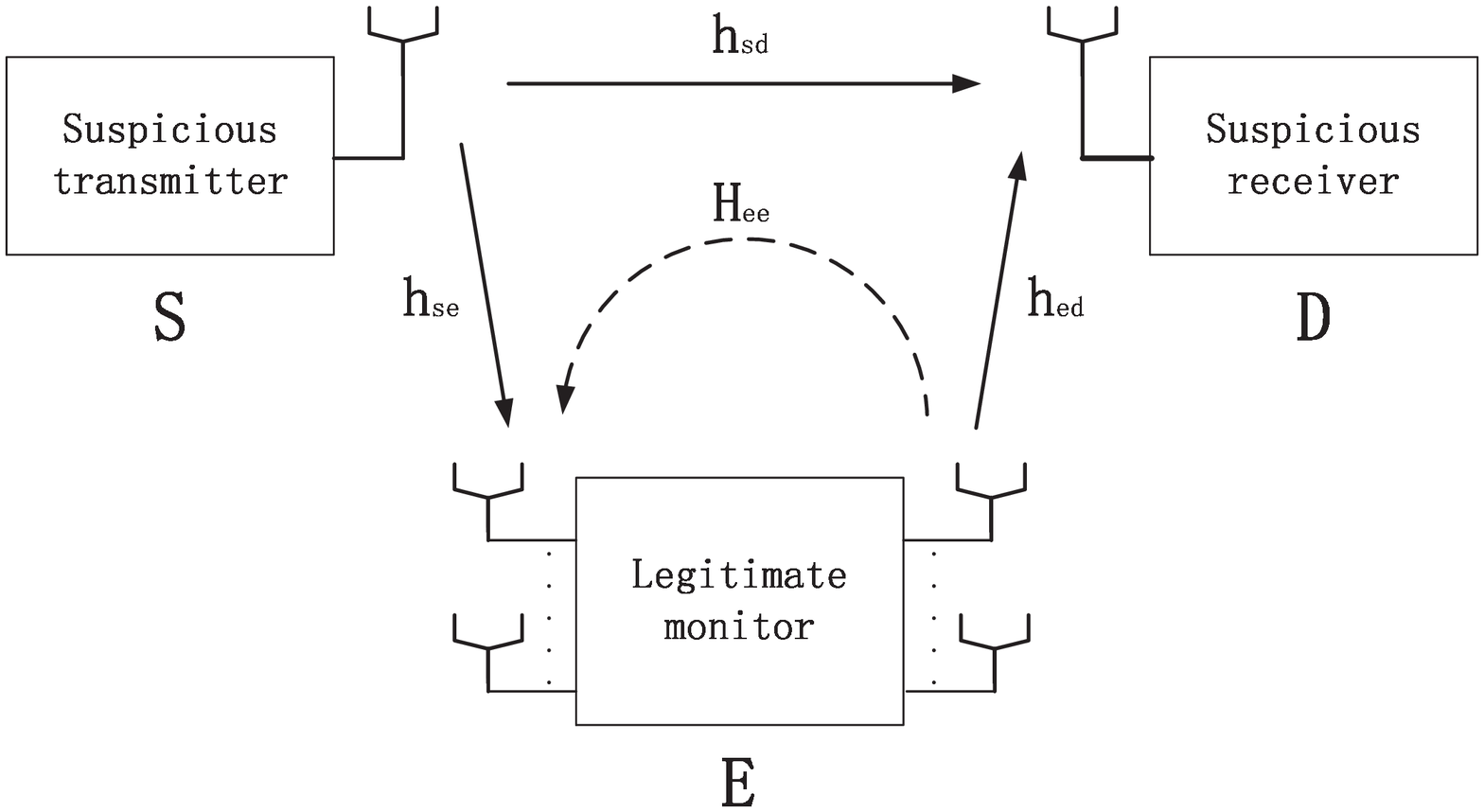}
\caption{A point-to-point legitimate surveillance system consisting of one suspicious transmitter S, one suspicious receiver D and one legitimate monitor E.}
\label{fig:fig1}
\end{figure}

We consider a three-node point-to-point legitimate surveillance system as shown in Fig. \ref{fig:fig1}, where a legitimate monitor E aims to eavesdrop a dubious communication link between a suspicious pair S and D via jamming. It is assumed that the suspicious transmitter and receiver are equipped with a single antenna each.\footnote{Although the current work focuses on the single antenna scenario, the developed approaches can be easily extended to the general multiple antenna scenario.} To enable simultaneous eavesdropping and jamming, the legitimate monitor is equipped with two sets of antennas, i.e., $N_r$ antennas for eavesdropping (receiving) and $N_t$ antennas for jamming (transmitting). Quasi-static channel fading is assumed, such that the channel coefficients remain unchanged during each transmission block but vary independently between different blocks.

The received signal at the suspicious receiver D can be expressed as
\begin{align} \label{SM:1}
y_D=\sqrt{P_S} h_{sd}s+ \mathbf{h}_{ed} \mathbf{w}_t x+n_d,
\end{align}
where $P_S$ denotes the transmit power of the suspicious transmitter, $h_{sd}$ is the channel coefficient of the $S \rightarrow D$ link which is a zero-mean complex Gaussian random variable with variance $\lambda_1$. The $1 \times N_t$ vector $\mathbf{h}_{ed}$ denotes the jamming channel between E and D, whose entries are identically and independently distributed (i.i.d.) zero-mean complex Gaussian random variables with variance $\lambda_3$ and $\mathbf{w}_t$ is the transmit beamforming vector at the legitimate monitor with $||\mathbf{w}_t||=1$. In addition, $s$ is the information symbol with unit power, while $x$ denotes the jamming symbol with ${\tt E}\{|x|^2\}=p_d$ satisfying $0 \leq p_d \leq P_J$ where $P_J$ denotes the maximum jamming power. Finally, $n_d$ is the zero-mean additive white Gaussian noise (AWGN) with variance $N_D$.

Similarly, the received signal at the legitimate monitor E is given by
\begin{align} \label{SM:2}
\mathbf{y}_E=\sqrt{P_S}  \mathbf{h}_{se}s+\sqrt{\rho}  \mathbf{H}_{ee} \mathbf{w}_t x+ \mathbf{n}_e,
\end{align}
where the $N_r \times 1$ vector $\mathbf{h}_{se}$ denotes the channel coefficient of the $S \rightarrow E$ link with entries being i.i.d. zero-mean complex Gaussian random variables with variance $\lambda_2$. As in \cite{G.Zheng,T.Riihonen}, the residual self-interference channel is modeled by $\sqrt{\rho} \mathbf{H}_{ee}$, where the $N_r \times N_t$ matrix $\mathbf{H}_{ee}$ denotes the fading loop channel with entries being i.i.d. zero-mean complex Gaussian random variables with variance $\lambda_4$ and $\rho$ ($0 \leq \rho \leq 1$) parameterizes the effect of passive self-interference suppression. Finally, $\mathbf{n}_e$ is the zero-mean AWGN noise at the legitimate monitor with ${\tt E} \{\mathbf{n}_e \mathbf{n}_e^\dagger\}=N_E \mathbf{I}_{N_r}$.

We assume that E employs a linear receiver $\mathbf{w}_r$ with $||\mathbf{w}_r||=1$ for signal detection, as such,
\begin{align} \label{SM:3}
y_E=\mathbf{w}_r^\dagger \mathbf{y}_E=\sqrt{P_S} \mathbf{w}_r^\dagger \mathbf{h}_{se}s+\sqrt{\rho} \mathbf{w}_r^\dagger \mathbf{H}_{ee} \mathbf{w}_t x+\mathbf{w}_r^\dagger \mathbf{n}_e.
\end{align}
Therefore, the end-to-end signal-to-interference-plus-noise ratio (SINR) at the suspicious receiver $\text{SINR}_D$ and the legitimate monitor $\text{SINR}_E$ can be respectively expressed as
\begin{align} \label{SM:4}
\text{SINR}_D=\frac{P_S|h_{sd}|^2}{p_d|\mathbf{h}_{ed} \mathbf{w}_t |^2+N_D} \quad \text{and} \quad \text{SINR}_E=\frac{P_S|\mathbf{w}_r^\dagger \mathbf{h}_{se}|^2}{\rho p_d|\mathbf{w}_r^\dagger \mathbf{H}_{ee} \mathbf{w}_t|^2+N_E}.
\end{align}

We assume that global channel state information (CSI) is available at the legitimate monitor \footnote{The CSI can be obtained by utilizing the methods given in the literature \cite{J.Xu1,Y.Zeng}.}, while the suspicious transmitter and receiver only know the CSI of the suspicious link. This assumption is practical since it is difficult for the suspicious transmitter to know the existence of the legitimate monitor. To ensure reliable detection at D, the suspicious transmitter varies the transmission rate according to $\text{SINR}_D$. Hence, if $\text{SINR}_E \geq \text{SINR}_D$, the legitimate monitor can also reliably decode the information. On the other hand, if $\text{SINR}_E < \text{SINR}_D$, it is impossible for the legitimate monitor to decode the information without any error. Therefore, we adopt the following indicator function to denote the event of successful eavesdropping at the legitimate monitor as in \cite{J.Xu1} :
\begin{align}  \label{SM:5}
X=\left\{\begin{array}{ll}
              1 & \text{if} \quad \text{SINR}_E \geq \text{SINR}_D,\\
             0 &  \text{otherwise},
           \end{array}\right.
\end{align}
where $X=1$ and $X=0$ denote eavesdropping non-outage and outage events, respectively. Note that the indicator function $X$ is irrespective of the transmit power $P_S$ at the suspicious transmitter.

As in \cite{J.Xu1}, we adopt the eavesdropping non-outage probability as the performance metric. Hence, the main objective is to maximize the eavesdropping non-outage probability ${\tt E}\{X\}$ by jointly optimizing the receive and transmit beamforming vector $\mathbf{w}_r$, $\mathbf{w}_t$ and the jamming power $p_d$. Hence, the optimization problem can be formulated as
\begin{align} \label{SM:6}
\text{(P1)} \quad : \quad \underset{ \mathbf{w}_r,\mathbf{w}_t,p_d}{\mathop{\max }} \quad & {\tt E}\{X\} \nonumber \\
\text{s.t.} \qquad & 0 \leq p_d \leq P_J \quad \& \quad ||\mathbf{w}_r||=||\mathbf{w}_t||=1.
\end{align}

\section{Optimal Design} \label{section:3}
In this section, we present optimal solutions for the optimization problem (P1). In particular, we investigate four different scenarios depending on the number of receive/transmit antennas implemented at the legitimate monitor. For each scenario, the optimization problem (P1) is reformulated and the optimal solution is obtained.

\subsection{Single-Input Single-Output (SISO)}
We start with the SISO case, which serves as a baseline scheme for comparison and as a useful guideline for the multiple-antenna cases. Since the legitimate monitor is equipped with a single transmit and receive antenna, i.e., $N_t=N_r=1$, self-interference can not be eliminated in the spatial domain but can be suppressed via proper jamming power design. As such, the optimization problem (P1) reduces to
\begin{align} \label{SISO:1}
\text{(P2)} \quad : \quad \underset{ p_d}{\mathop{\max }} \quad & {\tt E}\{X\}={\mathop{\rm Prob}\nolimits} \left( \frac{P_S| {h}_{se}|^2}{\rho p_d|{h}_{ee} |^2+N_E} \geq \frac{P_S|h_{sd}|^2}{p_d|{h}_{ed} |^2+N_D} \right) \nonumber \\
\text{s.t.} \quad & 0 \leq p_d \leq P_J .
\end{align}

Note that problem (P2) is non-convex in general, since its objective function is not concave over the jamming power $p_d$. However, problem (P2) can be reformulated as
\begin{align} \label{SISO:2}
\text{(P3)} \quad : \quad \underset{ p_d}{\mathop{\min }} \quad &  \frac{|h_{sd}|^2}{p_d|{h}_{ed} |^2+N_D}-\frac{| {h}_{se}|^2}{\rho p_d|{h}_{ee} |^2+N_E}   \nonumber \\
\text{s.t.} \quad & 0 \leq p_d \leq P_J .
\end{align}
Then we have the following result:
\begin{theorem}\label{theorem:1}
For the SISO scenario, the optimal jamming power can be expressed as
\begin{align}  \label{SISO:3}
p_d = \left\{
\begin{array}{ll}
P_J & \text{if} \quad \left\{ \begin{array}{ll} \Delta_1>0 &  \Delta_2>0 \quad \text{and} \quad  \frac{\Delta_2}{\Delta_1} \geq P_J \\  \Delta_1 = 0 &  \Delta_2 > 0 \\ \Delta_1 < 0 & \left\{ \begin{array}{ll} \Delta_2 \geq 0  \\ \Delta_2 < 0 \quad \text{and} \quad  \frac{\Delta_2}{\Delta_1} < P_J \quad \text{and} \quad \frac{\rho N_D |h_{se}|^2 |h_{ee}|^2}{\rho P_J|{h}_{ee}|^2 +N_E} \leq \frac{N_E|h_{sd}|^2|{h}_{ed}|^2}{P_J|{h}_{ed}|^2+N_D} \end{array} \right. \end{array} \right.\\
0 & \text{if} \quad \left\{ \begin{array}{ll} \Delta_1 \geq 0 & \Delta_2 \leq 0  \\ \Delta_1 < 0 & \Delta_2 < 0 \quad \left\{ \begin{array}{ll} \frac{\Delta_2}{\Delta_1} \geq P_J  \\ \frac{\Delta_2}{\Delta_1} < P_J \quad \text{and} \quad \frac{\rho N_D |h_{se}|^2 |h_{ee}|^2}{\rho P_J|{h}_{ee}|^2 +N_E} > \frac{N_E|h_{sd}|^2|{h}_{ed}|^2}{P_J|{h}_{ed}|^2+N_D} \end{array} \right. \end{array} \right. \\
 \frac{\Delta_2}{\Delta_1}  &  \text{if} \quad \Delta_1 > 0 \quad \text{and} \quad \Delta_2 >0 \quad \text{and} \quad \frac{\Delta_2}{\Delta_1} < P_J,
\end{array} \right.
\end{align}
where $\Delta_1=|h_{ed}|^2-\frac{\sqrt{\rho} |h_{sd}| |h_{ed}| |h_{ee}|}{|h_{se}|}$ and $\Delta_2=\frac{|h_{sd}| |h_{ed}|}{\sqrt{\rho} |h_{ee}| |h_{se}|} N_E-N_D$.
\proof  Define $f(x)=\frac{|h_{sd}|^2}{x|{h}_{ed} |^2+N_D}-\frac{| {h}_{se}|^2}{\rho x|{h}_{ee} |^2+N_E} $ with $ 0 \leq x \leq P_J$. The first order derivative of $f(x)$ can be expressed as
\begin{align} \label{SISO:4}
f'(x)=  -\frac{|h_{sd}|^2|{h}_{ed} |^2 }{\left(x|{h}_{ed} |^2+N_D \right)^2} + \frac{\rho |{h}_{ee}|^2 | {h}_{se}|^2}{\left(\rho x|{h}_{ee} |^2+N_E\right)^2}   .
\end{align}
To this end, the desired results can be obtained along with some separate treatments for different cases, for which the details are omitted for brevity.
\endproof
\end{theorem}

Note that the optimal jamming strategy depends on the relationship between the channel gains and noise powers. If jamming introduces higher level of interference power at the suspicious receiver than the self-interference power at the legitimate monitor, it is always beneficial to use full power to confuse the suspicious receiver. As for the scenario where the legitimate monitor can already overhear from the suspicious transmitter successfully without jamming or jamming causes higher self-interference at the legitimate monitor, it is better to remain silent. For some special scenarios that jamming causes non-monotonic influence on the eavesdropping performance, there exists an additional optimal power allocation point that achieves the best performance.

\subsection{Single-Input Multiple-Output (SIMO)}
We now consider the SIMO case. With multiple receive antennas, it is possible to mitigate the self-interference in the spatial domain, i.e., via the zero-forcing combining. However, complete elimination of the self-interference does not necessarily yield the highest effective SINR. Therefore, it is desirable to find the optimal receive combining vector. To do so, we first reformulate the original optimization problem (P1) as
\begin{align} \label{MISO:1}
\text{(P4)} \quad : \quad \underset{ p_d,\mathbf{w}_r}{\mathop{\max }} \quad & {\mathop{\rm Prob}\nolimits} \left( \frac{P_S|\mathbf{w}_r^\dagger \mathbf{h}_{se}|^2}{\rho p_d|\mathbf{w}_r^\dagger \mathbf{h}_{ee} |^2+N_E} \geq \frac{P_S|h_{sd}|^2}{p_d|{h}_{ed} |^2+N_D} \right) \nonumber \\
\text{s.t.} \quad \quad & 0 \leq p_d \leq P_J \quad \& \quad ||\mathbf{w}_r||=1.
\end{align}
To maximize the objective function of problem (P4), we find out that, for fixed $p_d$, it is sufficient to maximize $\frac{|\mathbf{w}_r^\dagger \mathbf{h}_{se}|^2}{\rho p_d|\mathbf{w}_r^\dagger \mathbf{h}_{ee} |^2+N_E}$. Since this is a generalized Rayleigh ratio problem \cite{R.A.Horn}, the optimal receiving vector can be obtained in closed-form as
\begin{align}   \label{MISO:2}
\mathbf{w}_r = \frac{ \left( \frac{\rho p_d}{N_E} \mathbf{h}_{ee} \mathbf{h}_{ee}^\dagger +\mathbf{I}_{N_r} \right)^{-1} \mathbf{h}_{se}}{||\left( \frac{\rho p_d}{N_E} \mathbf{h}_{ee} \mathbf{h}_{ee}^\dagger +\mathbf{I}_{N_r} \right)^{-1} \mathbf{h}_{se}||}.
\end{align}
Then, substituting $\mathbf{w}_r$ into (\ref{MISO:1}) and applying the Sherman Morrison formula \cite{W.W.Hager}, problem (P4) can be reformulated as
\begin{align}  \label{MISO:3}
\text{(P5)} \quad : \quad \underset{ p_d}{\mathop{\min }} \quad &  \frac{ N_E |h_{sd}|^2}{p_d|{h}_{ed}|^2+N_D} + \frac{\frac{\rho p_d}{N_E} |\mathbf{h}_{se}^\dagger \mathbf{h}_{ee} |^2}{1+\frac{\rho p_d}{N_E}||\mathbf{h}_{ee}||^2}  \nonumber \\
\text{s.t.} \quad &  0 \leq p_d \leq P_J.
\end{align}
Therefore, the remaining task is to find the optimal jamming power $p_d$, which we do in the following.

\begin{theorem}\label{theorem:2}
For the SIMO case, the optimal jamming power is given by
\begin{align}  \label{MISO:4}
p_d = \left\{
\begin{array}{ll}
P_J & \text{if} \quad \left\{ \begin{array}{ll} \Delta_1>0 &  \Delta_2>0 \quad \text{and} \quad  \frac{\Delta_2}{\Delta_1} \geq P_J \\  \Delta_1 = 0 &  \Delta_2 > 0 \\ \Delta_1 < 0 & \left\{ \begin{array}{ll} \Delta_2 \geq 0  \\ \Delta_2 < 0 \quad \text{and} \quad  \frac{\Delta_2}{\Delta_1} < P_J \quad \text{and} \quad \frac{\rho N_D |\mathbf{h}_{se}^\dagger \mathbf{h}_{ee} |^2}{\rho P_J||\mathbf{h}_{ee}||^2 +N_E} \leq \frac{N_E|h_{sd}|^2|{h}_{ed}|^2}{P_J|{h}_{ed}|^2+N_D} \end{array} \right. \end{array} \right.\\
0 & \text{if} \quad \left\{ \begin{array}{ll} \Delta_1 \geq 0 & \Delta_2 \leq 0  \\ \Delta_1 < 0 & \Delta_2 < 0 \quad \left\{ \begin{array}{ll} \frac{\Delta_2}{\Delta_1} \geq P_J  \\ \frac{\Delta_2}{\Delta_1} < P_J \quad \text{and} \quad \frac{\rho N_D |\mathbf{h}_{se}^\dagger \mathbf{h}_{ee} |^2}{\rho P_J||\mathbf{h}_{ee}||^2 +N_E} > \frac{N_E|h_{sd}|^2|{h}_{ed}|^2}{P_J|{h}_{ed}|^2+N_D} \end{array} \right. \end{array} \right. \\
 \frac{\Delta_2}{\Delta_1}  &  \text{if} \quad \Delta_1 > 0 \quad \text{and} \quad \Delta_2 >0 \quad \text{and} \quad \frac{\Delta_2}{\Delta_1} < P_J,
\end{array} \right.
\end{align}
where $\Delta_1=|h_{ed}|^2-\sqrt{\frac{\rho |h_{sd}|^2|h_{ed}|^2}{|\mathbf{h}_{se}^\dagger \mathbf{h}_{ee}|^2}}||\mathbf{h}_{ee}||^2$ and $\Delta_2=\sqrt{\frac{ |h_{sd}|^2|h_{ed}|^2}{\rho|\mathbf{h}_{se}^\dagger \mathbf{h}_{ee}|^2}}N_E-N_D$.
\proof  Define $g(x)=\frac{ N_E |h_{sd}|^2}{x |{h}_{ed}|^2+N_D} + \frac{\frac{\rho x}{N_E} |\mathbf{h}_{se}^\dagger \mathbf{h}_{ee} |^2}{1+\frac{\rho x}{N_E}||\mathbf{h}_{ee}||^2}$ with $ 0 \leq x \leq P_J$. The first order derivative of $g(x)$ can be computed as
\begin{align} \label{MISO:5}
g'(x)= - \frac{ N_E |h_{sd}|^2|{h}_{ed}|^2}{\left(x |{h}_{ed}|^2+N_D\right)^2} + \frac{\frac{\rho }{N_E} |\mathbf{h}_{se}^\dagger \mathbf{h}_{ee} |^2}{\left(1+\frac{\rho x}{N_E}||\mathbf{h}_{ee}||^2\right)^2}.
\end{align}
To this end, the desired results can be obtained along with some separate treatments for different cases, for which the details are omitted for brevity.
\endproof
\end{theorem}

Note that for the special case $N_r=1$, Theorem \ref{theorem:2} reduces to Theorem \ref{theorem:1}.

\subsection{Multiple-input Single-output (MISO)}
In this subsection, we focus on the MISO case. Different from the SIMO case, where the receive vector design only affects the effective $\text{SINR}_E$, the transmit beamforming vector will affect both $\text{SINR}_E$ and $\text{SINR}_D$, hence it is more challenging to design.

To start with, substituting (\ref{SM:4}) into (\ref{SM:6}), problem (P1) can be alternatively expressed as
\begin{align} \label{SIMO:1}
\text{(P6)} \quad : \quad \underset{ p_d,\mathbf{w}_t}{\mathop{\max }} \quad & {\mathop{\rm Prob}\nolimits} \left( \frac{P_S|{h}_{se}|^2}{\rho p_d|\mathbf{h}_{ee} \mathbf{w}_t|^2+N_E} \geq \frac{P_S|h_{sd}|^2}{p_d|\mathbf{h}_{ed} \mathbf{w}_t |^2+N_D} \right) \nonumber \\
\text{s.t.} \quad \quad & 0 \leq p_d \leq P_J \quad \& \quad ||\mathbf{w}_t||=1.
\end{align}
With simple algebraic manipulations, problem (P6) can be equivalently formulated as
\begin{align}  \label{SIMO:2}
\text{(P7)} \quad : \quad \underset{ p_d,\mathbf{w}_t}{\mathop{\min }} \quad &  \frac{\rho p_d|\mathbf{h}_{ee} \mathbf{w}_t|^2+N_E}{p_d|\mathbf{h}_{ed} \mathbf{w}_t |^2+N_D} \nonumber \\
\text{s.t.} \quad \quad & 0 \leq p_d \leq P_J \quad \& \quad ||\mathbf{w}_t||=1.
\end{align}
To this end, we have the following important observation:
\begin{lemma}\label{lemma:1}
For the MISO case, it is always optimal to use the full jamming power, i.e., $p_d=P_J$.
\proof Capitalizing on the technique presented in \cite[Lemma 4]{G.Zheng}, the claim can be established. \endproof
\end{lemma}

Therefore, the remaining task is to optimize $\mathbf{w}_t$. Let $\mathbf{W}\triangleq\mathbf{w}_t\mathbf{w}_t^\dagger$ and ignore the rank-one constraint for the moment, then problem (P7) can be relaxed as:
\begin{align} \label{SIMO:4}
\text{(P8)} \quad : \quad \underset{ \mathbf{W}}{\mathop{\min }} \quad &   \frac{\rho P_J \text{tr}\left(\mathbf{W} \mathbf{h}_{ee}^\dagger  \mathbf{h}_{ee} \right)+N_E}{P_J \text{tr}\left( \mathbf{W} \mathbf{h}_{ed}^\dagger \mathbf{h}_{ed}\right)+N_D}  \\
\text{s.t.} \quad
& \text{tr} \left(\mathbf{W}\right) = 1  \tag{18a} \\
& \mathbf{W} \succeq \mathbf{0}. \tag{18b}
\end{align}

It is easy to observe that problem (P8) is a fractional semidefinite programming (SDP), which can be converted into an equivalent linear SDP through the Charnes-Cooper transformation \cite{A.Charnes}. Specifically, introducing a new variable $\mathbf{Z}=s\mathbf{W}$, where $s>0$ satisfies $s\left(P_J \text{tr}\left( \mathbf{W} \mathbf{h}_{ed}^\dagger \mathbf{h}_{ed}\right)+N_D\right)=1$. Then multiplying the numerator and the denominator of the objective function by $s$, the fractional SDR of problem (P8) becomes a convex SDP as follows
\begin{align} \label{SIMO:5}
\text{(P9)} \quad : \quad \underset{ \mathbf{Z},s}{\mathop{\min }} \quad &  \rho P_J \text{tr}\left(\mathbf{Z} \mathbf{h}_{ee}^\dagger  \mathbf{h}_{ee} \right)+sN_E  \\
\text{s.t.} \quad & s > 0 \tag{19a} \\
& \text{tr} \left(\mathbf{Z}\right) = s  \tag{19b} \\
&  P_J \text{tr}\left( \mathbf{Z} \mathbf{h}_{ed}^\dagger \mathbf{h}_{ed}\right)+sN_D=1 \tag{19c} \\
& \mathbf{Z} \succeq \mathbf{0}. \tag{19d}
\end{align}

Problem (P9) is a convex SDP problem that consists of a linear objective function with a set of linear constraints. Therefore, the optimal solution can be efficiently solved using the standard CVX tools \cite{S.Boyd}. Then, we have the following lemma:
\begin{lemma}\label{lemma:2}
The fractional quasi-convex problem (P8) attains the same optimal objective value as that of problem (P9). Furthermore, if $\left(\mathbf{Z}^*,s^*\right)$ is the optimal solution of (P9), then $\frac{\mathbf{Z}^*}{s^*}$ is the optimal solution of (P8).
\proof The proof can be established by using a similar technique proposed in \cite{Z.Zhang,Q.Li}. Hence, it is omitted for brevity.  \endproof
\end{lemma}

Recall that the rank-one constraint has been neglected in problem (P9). Hence, to establish the optimality of the solution, we need to verify the rank of the optimum $\mathbf{Z}$, which we show in the following theorem:
\begin{theorem}\label{theorem:3}
The optimal solution $\mathbf{Z}$ of problem (P9) is always rank one.
\proof  See Appendix \ref{proof:theorem:3}.  \endproof
\end{theorem}

\subsection{Multiple-input Multiple-output (MIMO)}
Now we turn our attentions to the most general MIMO case. Substituting (\ref{SM:4}) into (\ref{SM:6}), problem (P1) can be alternatively expressed as
\begin{align} \label{MIMO:1}
\text{(P10)} \quad : \quad \underset{ p_d,\mathbf{w}_r,\mathbf{w}_t}{\mathop{\max }} \quad & {\mathop{\rm Prob}\nolimits} \left( \frac{P_S|\mathbf{w}_r^\dagger \mathbf{h}_{se}|^2}{\rho p_d|\mathbf{w}_r^\dagger \mathbf{H}_{ee} \mathbf{w}_t|^2+N_E} \geq \frac{P_S|h_{sd}|^2}{p_d|\mathbf{h}_{ed} \mathbf{w}_t |^2+N_D} \right) \nonumber \\
\text{s.t.} \quad \quad & 0 \leq p_d \leq P_J \quad \& \quad ||\mathbf{w}_r||=||\mathbf{w}_t||=1.
\end{align}
To proceed, we rewrite problem (P10) as
\begin{align} \label{MIMO:2}
\text{(P11)} \quad : \quad \underset{ p_d,\mathbf{w}_r,\mathbf{w}_t}{\mathop{\min }} \quad &  \frac{|h_{sd}|^2}{p_d|\mathbf{h}_{ed} \mathbf{w}_t |^2+N_D}- \frac{|\mathbf{w}_r^\dagger \mathbf{h}_{se}|^2}{\rho p_d|\mathbf{w}_r^\dagger \mathbf{H}_{ee} \mathbf{w}_t|^2+N_E} \nonumber \\
\text{s.t.} \quad \quad & 0 \leq p_d \leq P_J \quad \& \quad ||\mathbf{w}_r||=||\mathbf{w}_t||=1.
\end{align}
Then, following the same argument as Lemma \ref{lemma:1}, it can be shown that using full jamming power is always optimal. Similarly, noticing that the item $\frac{|\mathbf{w}_r^\dagger \mathbf{h}_{se}|^2}{\rho p_d|\mathbf{w}_r^\dagger \mathbf{H}_{ee} \mathbf{w}_t|^2+N_E}$ in (\ref{MIMO:2}) is a generalized Rayleigh ratio problem, which can be globally maximized when
\begin{align} \label{MIMO:3}
\mathbf{w}_r = \frac{ \left( \frac{\rho p_d}{N_E} \mathbf{H}_{ee} \mathbf{w}_t \mathbf{w}_t^\dagger \mathbf{H}_{ee}^\dagger +\mathbf{I}_{N_r} \right)^{-1} \mathbf{h}_{se}}{||\left( \frac{\rho p_d}{N_E} \mathbf{H}_{ee} \mathbf{w}_t \mathbf{w}_t^\dagger \mathbf{H}_{ee}^\dagger +\mathbf{I}_{N_r} \right)^{-1} \mathbf{h}_{se}||}.
\end{align}
Hence, problem (P11) can be reformulated as
\begin{align} \label{MIMO:4}
\text{(P12)} \quad : \quad \underset{ \mathbf{w}_t}{\mathop{\min }} \quad &  \frac{ \frac{N_E}{N_D}|h_{sd}|^2}{1+\frac{P_J}{N_D}|\mathbf{h}_{ed} \mathbf{w}_t|^2} + \frac{\frac{\rho P_J}{N_E} |\mathbf{h}_{se}^\dagger \mathbf{H}_{ee} \mathbf{w}_t|^2}{1+\frac{\rho P_J}{N_E}\mathbf{w}_t^\dagger \mathbf{H}_{ee}^\dagger \mathbf{H}_{ee} \mathbf{w}_t}  \nonumber \\
\text{s.t.} \quad &  ||\mathbf{w}_t||=1.
\end{align}
The optimization problem (P12) is non-convex because of the complex objective function. To proceed, we first introduce an auxiliary variable $y$ as
\begin{align}  \label{MIMO:5}
y=1+\frac{P_J}{N_D}|\mathbf{h}_{ed} \mathbf{w}_t|^2=1+\frac{P_J}{N_D} \text{tr}\left(\mathbf{W} \mathbf{h}_{ed}^\dagger \mathbf{h}_{ed}\right).
\end{align}
Apparently, $y\in \left[1,y_{\text{max}}\right]$, where $y_{\text{max}}$ is the maximum eigenvalue of matrix $\mathbf{I}_{N_t}+\frac{P_J}{N_D} \mathbf{h}_{ed}^\dagger \mathbf{h}_{ed}$, i.e., $y_{\text{max}}=1+\frac{P_J}{N_D} ||\mathbf{h}_{ed}||^2$.

Then, employing the SDR technique, problem (P12) can be relaxed as
\begin{align} \label{MIMO:6}
\text{(P13)} \quad : \quad \underset{ \mathbf{W}}{\mathop{\min }} \quad &   \frac{\frac{\rho P_J}{N_E} \text{tr}\left(\mathbf{W} \mathbf{H}_{ee}^\dagger \mathbf{h}_{se} \mathbf{h}_{se}^\dagger \mathbf{H}_{ee} \right)}{1+\frac{\rho P_J}{N_E} \text{tr}\left( \mathbf{W} \mathbf{H}_{ee}^\dagger \mathbf{H}_{ee}\right)}+ \frac{ \frac{N_E}{N_D}|h_{sd}|^2}{y}   \\
\text{s.t.} \quad &  y = 1+\frac{P_J}{N_D}\text{tr} \left(\mathbf{W} \mathbf{h}_{ed}^\dagger \mathbf{h}_{ed}\right) \tag{25a} \\
& \text{tr} \left(\mathbf{W}\right) = 1  \tag{25b} \\
& \mathbf{W} \succeq \mathbf{0}, \tag{25c}
\end{align}
where the rank-one constraint is omitted. Note that when $y$ is fixed, problem (P13) becomes a quasi-convex optimization problem, which can be converted into a convex SDP problem after some transformations. Hence, problem (P13) can be solved by the two-stage optimization procedure \cite{J.Zhou}, where the inner stage is a SDP problem with fixed $y$, while the outer stage is a one dimensional line search problem over $y$.

In particular, the one dimensional problem is formulated by
\begin{align} \label{MIMO:7}
\text{(P14)} \quad : \quad \underset{ y}{\mathop{\min }} \quad &   f(y)+ \frac{ \frac{N_E}{N_D}|h_{sd}|^2}{y}  \nonumber \\
\text{s.t.} \quad & 1 \leq y \leq 1+\frac{P_J}{N_D} ||\mathbf{h}_{ed}||^2,
\end{align}
where $f(y)$ is the optimal value of the inner optimization problem (P15) presented below:
\begin{align} \label{MIMO:8}
\text{(P15)} \quad : \quad \underset{ \mathbf{Z},s}{\mathop{\min }} \quad &   \frac{\rho P_J}{N_E} \text{tr}\left(\mathbf{Z} \mathbf{H}_{ee}^\dagger \mathbf{h}_{se} \mathbf{h}_{se}^\dagger \mathbf{H}_{ee} \right)   \\
\text{s.t.} \quad & s>0   \tag{27a} \\
& \text{tr} \left(\mathbf{Z}\right) = s  \tag{27b} \\
& s+\frac{\rho P_J}{N_E} \text{tr} \left(\mathbf{Z \mathbf{H}_{ee}^\dagger \mathbf{H}_{ee}}\right)=1  \tag{27c} \\
& \frac{P_J}{N_D}\text{tr} \left(\mathbf{Z} \mathbf{h}_{ed}^\dagger \mathbf{h}_{ed}\right) = s(y-1) \tag{27d} \\
& \mathbf{Z} \succeq \mathbf{0}, \tag{27e}
\end{align}
where we have used the same technique as in problem (P8) by introducing a new variable $\mathbf{Z}=s\mathbf{W}$, where $s>0$ satisfies $s+\frac{\rho P_J}{N_E}\text{tr} \left(s\mathbf{W} \mathbf{H}_{ee}^\dagger \mathbf{H}_{ee}\right)=1$.

Problem (P15) consists of a linear objective function with a set of linear constraints, hence is a convex SDP problem that can be efficiently solved. Recall that the SDR technique was applied to facilitate the derivation. Hence, it remains to check whether the solution of problem (P15) satisfies the rank-one constraint. Then we have the following important result:
\begin{theorem}\label{theorem:4}
The optimal $\mathbf{Z}$ of problem (P15) is guaranteed to be rank-one.
\proof  See Appendix \ref{proof:theorem:4}.  \endproof
\end{theorem}

\section{Suboptimal Design for the MIMO Case and Performance Analysis}  \label{section:4}
In the previous section, we have studied the optimal design for the MIMO case. However, the resulting solution requires one dimensional search and SDP, hence involves high computation complexity. Motivated by this, in this section, we propose three low-complexity suboptimal beamforming design schemes. In addition, a detailed analysis on the exact eavesdropping non-outage probability of the corresponding systems is presented.

\subsection{SISO Case}
We start by investigating the achievable performance of the SISO case, which serves as a baseline scheme for comparison, and we have the following result:
\begin{theorem}\label{theorem:5}
For the SISO case, the exact eavesdropping non-outage probability of the system is given as
\begin{multline} \label{PA:1}
{\tt E}\{X\}=1+\left(\frac{\lambda_1N_E}{P_J(\rho\lambda_1\lambda_4-\lambda_2\lambda_3)}-\frac{\rho \lambda_1\lambda_2\lambda_3\lambda_4}{\left(\rho\lambda_1\lambda_4-\lambda_2\lambda_3\right)^2}\right) \exp \left(\frac{N_D}{\lambda_3 P_J}+\frac{\lambda_1N_E}{\lambda_2 \lambda_3 P_J}\right) \Gamma \left(0,\frac{N_D}{\lambda_3 P_J}+\frac{\lambda_1N_E}{\lambda_2 \lambda_3 P_J}\right) \\ - \frac{\rho \lambda_1^2 \lambda_4 N_E}{(\rho \lambda_1\lambda_4-\lambda_2\lambda_3)(\lambda_1N_E+\lambda_2N_D)}+\frac{\rho \lambda_1\lambda_2\lambda_3\lambda_4}{\left(\rho\lambda_1\lambda_4-\lambda_2\lambda_3\right)^2} \exp \left(\frac{N_D}{\lambda_3 P_J}+\frac{N_E}{\rho \lambda_4 P_J}\right)\Gamma \left(0,\frac{N_D}{\lambda_3 P_J}+\frac{N_E}{\rho \lambda_4 P_J}\right) .
\end{multline}
\proof  See Appendix \ref{proof:theorem:5}.  \endproof
\end{theorem}

Theorem \ref{theorem:5} presents a closed-form expression for the eavesdropping non-outage probability, which is valid for arbitrary system configuration. Nevertheless, the expression does not provide much insightful information. Motivated by this, we look into the high SNR regime and derive an asymptotic approximation for the system, which enables the characterization of the achievable diversity order.

In conventional physical layer security literatures, a common assumption in the asymptotic high SNR regime is that the main-to-eavesdropper ratio (MER) $\lambda_{de} \rightarrow \infty$ (i.e., the ratio between the reference gains of the main channel and eavesdropping channel), see for instance \cite{Y.Zou,Y.Zou1,L.Fan,J.Zhu}. In practice, this occurs when the quality of the main channel is much better than wiretap channel, i.e., Bob is relatively close to Alice while Eve is far away from Alice or the wiretap channel undergoes severe small-scale and large-scale fading effects. Similarly, we propose a new metric, namely, eavesdropper-to-main ratio (EMR) as $\lambda_{ed}=\lambda_2/\lambda_1$, and define the diversity order of the system as
\begin{align}  \label{PA:2}
d_{\text{EMR
}} =-  \lim\limits_{\lambda_{ed} \rightarrow \infty} \frac{\log \left(P_{\sf out}^\infty\right)}{\log \left(\lambda_{ed}\right)}.
\end{align}

\begin{lemma} \label{lemma:3}
In the high SNR regime, i.e., $\lambda_{ed} \rightarrow \infty$, the eavesdropping outage probability of the SISO case can be approximated as
\begin{multline}  \label{PA:3}
P_{\sf out}^{\infty}=\left(\left(\frac{\rho \lambda_4}{\lambda_3}+\frac{N_E}{\lambda_3 P_J}\right)\exp \left(\frac{N_D}{\lambda_3P_J}\right) \Gamma\left(0,\frac{N_D}{\lambda_3 P_J}\right)- \right. \\ \left. \frac{\rho \lambda_4}{\lambda_3} \exp \left(\frac{N_D}{\lambda_3 P_J}+\frac{N_E}{\rho \lambda_4 P_J}\right)\Gamma\left(0,\frac{N_D}{\lambda_3 P_J}+\frac{N_E}{\rho \lambda_4 P_J}\right)\right)\times \frac{1}{\lambda_{ed}}.
\end{multline}
\proof  See Appendix \ref{proof:lemma:3}.  \endproof
\end{lemma}

Lemma \ref{lemma:3} indicates that the system achieves a unit diversity order. This is intuitive since only one receive antenna is deployed at the legitimate monitor.

\subsection{TZF/MRC Scheme}
The basic idea of the TZF/MRC scheme is to exploit the multiple transmit antennas to completely eliminate the self-interference \cite{H.A.Suraweera}. To ensure this is feasible, the number of the transmit antennas should be greater than one, i.e., $N_t > 1$. In addition, MRC is applied at the receive antennas, i.e.,
\begin{align}  \label{sub:1}
\mathbf{w}_r = \frac{\mathbf{h}_{se}}{||\mathbf{h}_{se}||}.
\end{align}
Hence, the optimal beamforming vector $\mathbf{w}_t$ is the solution of the following problem
\begin{align} \label{sub:2}
&\mathbf{w}_t= \arg \max\limits_{\mathbf{w}_t} |\mathbf{h}_{ed} \mathbf{w}_t|^2 \nonumber \\
&\text{s.t.} \quad \mathbf{h}_{se}^\dagger \mathbf{H}_{ee} \mathbf{w}_t=0 \quad \& \quad ||\mathbf{w}_t||=1.
\end{align}
According to \cite{M.Mohammadi}, the optimal solution can be written in closed-form as
\begin{align} \label{sub:3}
\mathbf{w}_t=\frac{\mathbf{\Pi_1} \mathbf{h}_{ed}^\dagger}{||\mathbf{\Pi_1} \mathbf{h}_{ed}^\dagger||},
\end{align}
where the $N_t \times N_t$ matrix $\mathbf{\Pi_1}$ is defined by $\mathbf{\Pi_1} = \mathbf{I}_{N_t}-\frac{\mathbf{H}_{ee}^\dagger \mathbf{h}_{se}\mathbf{h}_{se}^\dagger \mathbf{H}_{ee}}{||\mathbf{h}_{se}^\dagger \mathbf{H}_{ee}||^2}$. Then we have the following result:
\begin{theorem}\label{theorem:6}
The exact eavesdropping non-outage probability of the TZF/MRC scheme can be expressed as
\begin{multline}  \label{PA:4}
  {\tt E}\{X\} = 1-\sum_{k=0}^{N_t-2} \frac{(-1)^k}{k! (N_t-k-2)!} \left(\frac{\lambda_1 N_E}{\lambda_2 \lambda_3 P_J}\right)^{N_r} \times \\ \left(\frac{N_D}{\lambda_3 P_J}+\frac{\lambda_1 N_E}{\lambda_2 \lambda_3 P_J} \right)^k \exp \left(\frac{N_D}{\lambda_3 P_J}+\frac{\lambda_1 N_E}{\lambda_2 \lambda_3 P_J} \right) \Gamma \left(N_t-N_r-k-1,\frac{N_D}{\lambda_3 P_J}+\frac{\lambda_1 N_E}{\lambda_2 \lambda_3 P_J} \right).
\end{multline}
\proof  See Appendix \ref{proof:theorem:6}.  \endproof
\end{theorem}

To gain further insights, we now look into the high SNR regime.

\begin{lemma} \label{lemma:4}
In the high SNR regime, i.e., $\lambda_{ed} \rightarrow \infty$, the eavesdropping outage probability of the TZF/MRC scheme can be approximated as
\begin{align} \label{PA:5}
P_{\sf out}^{\infty}=\sum_{k=0}^{N_t-2} \frac{(-1)^k}{k!(N_t-k-2)!} \left(\frac{N_D}{\lambda_3 P_J}\right)^k e^{\frac{N_D}{\lambda_3 P_J}} \Gamma \left(N_t-N_r-k-1,\frac{N_D}{\lambda_3 P_J}\right) \left(\frac{N_E}{ \lambda_3 P_J}\right)^{N_r} \times \left(\frac{1}{\lambda_{ed}}\right)^{N_r}.
\end{align}
\proof  See Appendix \ref{proof:lemma:4}.  \endproof
\end{lemma}

From Lemma \ref{lemma:4}, we observe that the system achieves a full diversity order of $N_r$, indicating that increasing the receive antenna number is an effective means to improve the system performance.

\subsection{MRT/RZF Scheme}
In contrast to the TZF/MRC scheme, self-interference cancellation is performed at the receiver by using RZF. To ensure this is feasible, the number of the receive antennas should be greater than one, i.e., $N_r > 1$. In addition, MRT is applied at the transmit antennas, i.e.,
\begin{align} \label{sub:4}
\mathbf{w}_t = \frac{\mathbf{h}_{ed}^\dagger}{||\mathbf{h}_{ed}||}.
\end{align}
Hence, the optimal beamforming vector $\mathbf{w}_r$ is the solution of the following maximization problem
\begin{align}  \label{sub:5}
&\mathbf{w}_r= \arg \max\limits_{\mathbf{w}_r} |\mathbf{w}_r^\dagger \mathbf{h}_{se}|^2 \nonumber \\
&\text{s.t.} \quad \mathbf{w}_r^\dagger \mathbf{H}_{ee}\mathbf{h}_{ed}^\dagger =0 \quad \& \quad ||\mathbf{w}_r||=1.
\end{align}
According to \cite{M.Mohammadi}, the optimal solution can be written in closed-form as
\begin{align} \label{sub:6}
\mathbf{w}_r=\frac{\mathbf{\Pi_2} \mathbf{h}_{se}}{||\mathbf{\Pi_2} \mathbf{h}_{se}||},
\end{align}
where the $N_r \times N_r$ matrix $\mathbf{\Pi_2}$ is given by $\mathbf{\Pi_2} = \mathbf{I}_{N_r}-\frac{\mathbf{H}_{ee} \mathbf{h}_{ed}^\dagger \mathbf{h}_{ed} \mathbf{H}_{ee}^\dagger }{|| \mathbf{H}_{ee} \mathbf{h}_{ed}^\dagger||^2}$. Then we have the following result:
\begin{theorem}\label{theorem:7}
The exact eavesdropping non-outage probability of the MRT/RZF scheme can be expressed as
\begin{multline}  \label{PA:6}
  {\tt E}\{X\} = 1-\sum_{k=0}^{N_t-1} \frac{(-1)^k}{k! (N_t-k-1)!} \left(\frac{\lambda_1 N_E}{\lambda_2 \lambda_3 P_J}\right)^{N_r-1} \times \\ \left(\frac{N_D}{\lambda_3 P_J}+\frac{\lambda_1 N_E}{\lambda_2 \lambda_3 P_J} \right)^k \exp \left(\frac{N_D}{\lambda_3 P_J}+\frac{\lambda_1 N_E}{\lambda_2 \lambda_3 P_J} \right) \Gamma \left(N_t-N_r-k+1,\frac{N_D}{\lambda_3 P_J}+\frac{\lambda_1 N_E}{\lambda_2 \lambda_3 P_J} \right).
\end{multline}
\proof  The desired result can be obtained by following similar procedures as in Appendix \ref{proof:theorem:6}.  \endproof
\end{theorem}

Since the above expression is too complicated to gain more insights, we now study the high SNR regime.

\begin{lemma} \label{lemma:5}
In the high SNR regime, i.e., $\lambda_{ed} \rightarrow \infty$, the eavesdropping outage probability of the MRT/RZF scheme can be approximated as
\begin{align} \label{PA:7}
P_{\sf out}^{\infty}=\sum_{k=0}^{N_t-1} \frac{(-1)^k}{k!(N_t-k-1)!} \left(\frac{N_D}{\lambda_3 P_J}\right)^k e^{\frac{N_D}{\lambda_3 P_J}} \Gamma \left(N_t-N_r-k+1,\frac{N_D}{\lambda_3 P_J}\right) \left(\frac{ N_E}{\lambda_3 P_J}\right)^{N_r-1} \times \left(\frac{1}{\lambda_{ed}}\right)^{N_r-1}.
\end{align}
\proof  The desired result can be obtained by following similar procedures as in Appendix \ref{proof:lemma:4}.  \endproof
\end{lemma}

Lemma \ref{lemma:5} reveals that the system achieves a diversity order of $N_r-1$. The reason is that one degree of freedom is used for interference cancellation at the receive side of the legitimate monitor.

\subsection{MRT/MRC Scheme}
Finally, we consider the MRT/MRC scheme. Hence, the beamforming vectors are given by
\begin{align}  \label{sub:7}
\mathbf{w}_r= \frac{\mathbf{h}_{se}}{||\mathbf{h}_{se}||} \quad \text{and} \quad \mathbf{w}_t = \frac{\mathbf{h}_{ed}^\dagger}{||\mathbf{h}_{ed}||}.
\end{align}

It is worth pointing out that, unlike the ZF schemes, the jamming power $p_d$ needs to be optimized due to the existence of self-interference in the MRT/MRC scheme. By following the same steps as in Theorem \ref{theorem:1}, we have the following result:

\begin{theorem}\label{theorem:8}
The optimal jamming power design for the MRT/MRC scheme can be expressed as
\begin{align}  \label{sub:8}
p_d = \left\{
\begin{array}{ll}
P_J & \text{if} \quad \left\{ \begin{array}{ll} \Delta_1>0 &  \Delta_2>0 \quad \text{and} \quad  \frac{\Delta_2}{\Delta_1} \geq P_J \\  \Delta_1 = 0 &  \Delta_2 > 0 \\ \Delta_1 < 0 & \left\{ \begin{array}{ll} \Delta_2 \geq 0  \\ \Delta_2 < 0 \quad \text{and} \quad  \frac{\Delta_2}{\Delta_1} < P_J \quad \text{and} \quad \frac{\rho N_D  |\mathbf{h}_{se}^\dagger \mathbf{H}_{ee} \mathbf{h}_{ed}^\dagger|^2}{\rho P_J \frac{|\mathbf{h}_{se}^\dagger \mathbf{H}_{ee} \mathbf{h}_{ed}^\dagger|^2}{||\mathbf{h}_{se}||^2} +N_E ||\mathbf{h}_{ed}||^2 } \leq \frac{N_E|h_{sd}|^2||\mathbf{h}_{ed}||^2}{P_J||\mathbf{h}_{ed}||^2+N_D} \end{array} \right. \end{array} \right.\\
0 & \text{if} \quad \left\{ \begin{array}{ll} \Delta_1 \geq 0 & \Delta_2 \leq 0  \\ \Delta_1 < 0 & \Delta_2 < 0 \quad \left\{ \begin{array}{ll} \frac{\Delta_2}{\Delta_1} \geq P_J  \\ \frac{\Delta_2}{\Delta_1} < P_J \quad \text{and} \quad \frac{\rho N_D  |\mathbf{h}_{se}^\dagger \mathbf{H}_{ee} \mathbf{h}_{ed}^\dagger|^2}{\rho P_J \frac{|\mathbf{h}_{se}^\dagger \mathbf{H}_{ee} \mathbf{h}_{ed}^\dagger|^2}{||\mathbf{h}_{se}||^2} +N_E ||\mathbf{h}_{ed}||^2 } > \frac{N_E|h_{sd}|^2||\mathbf{h}_{ed}||^2}{P_J||\mathbf{h}_{ed}||^2+N_D} \end{array} \right. \end{array} \right. \\
 \frac{\Delta_2}{\Delta_1}  &  \text{if} \quad \Delta_1 > 0 \quad \text{and} \quad \Delta_2 >0 \quad \text{and} \quad \frac{\Delta_2}{\Delta_1} < P_J,
\end{array} \right.
\end{align}
where $\Delta_1=||\mathbf{h}_{ed}||^2-\frac{\sqrt{\rho} |h_{sd}|  |\mathbf{h}_{se}^\dagger \mathbf{H}_{ee} \mathbf{h}_{ed}^\dagger|}{||\mathbf{h}_{se}||^2}$ and $\Delta_2=\frac{|h_{sd}| ||\mathbf{h}_{ed}||^2}{\sqrt{\rho} |\mathbf{h}_{se}^\dagger \mathbf{H}_{ee} \mathbf{h}_{ed}^\dagger|} N_E-N_D$.
\end{theorem}

Having obtained the optimal jamming power, we are ready to study the exact eavesdropping non-outage probability of MRT/MRC scheme.
\begin{theorem}\label{theorem:9}
The exact eavesdropping non-outage probability of the MRT/MRC scheme is given by
\begin{multline}  \label{PA:8}
  {\tt E}\{X\} = 1-\frac{1}{\left(1+\frac{\lambda_2 N_D}{\lambda_1 N_E}\right)^{N_r}}+ \\ \sum_{k=0}^{N_t-1} \sum_{m=0}^k \frac{1}{(k-m)!} \int_{\frac{N_D}{N_E}}^\infty \left(\frac{x N_E-N_D}{\lambda_3P_J}\right)^{k-m} e^{-\frac{x N_E-N_D}{\lambda_3P_J}}
\frac{\left(\frac{\rho\lambda_4}{\lambda_3}x\right)^m}{\left(1+\frac{\rho \lambda_4}{\lambda_3}x\right)^{m+1}} \frac{N_r \frac{\lambda_2}{\lambda_1}}{\left(1+\frac{\lambda_2}{\lambda_1}x\right)^{N_r+1}} \rm dx.
\end{multline}
\proof  See Appendix \ref{proof:theorem:9}.  \endproof
\end{theorem}

To the best of the authors' knowledge, the integral in (\ref{PA:8}) does not admit a closed-form expression. However, it can be efficiently evaluated numerically using standard software such as Matlab or Mathematica. To gain further insights, we now look into the high SNR regime.

\begin{lemma} \label{lemma:6}
In the high SNR regime, i.e., $\lambda_{ed} \rightarrow \infty$, the eavesdropping outage probability of the MRT/MRC scheme can be approximated as
\begin{multline} \label{PA:9}
P_{\sf out}^{\infty}=\left(\left(\frac{N_E}{N_D}\right)^{N_r}-\sum_{k=0}^{N_t-1} \sum_{m=0}^k \frac{N_r}{(k-m)!} \int_{\frac{N_D}{N_E}}^\infty \left(\frac{x N_E-N_D}{\lambda_3P_J}\right)^{k-m} e^{-\frac{x N_E-N_D}{\lambda_3P_J}}
\frac{\left(\frac{\rho\lambda_4}{\lambda_3}\right)^m x^{m-N_r-1}}{\left(1+\frac{\rho \lambda_4}{\lambda_3}x\right)^{m+1}}  {\rm dx} \right) \\ \times \left(\frac{1}{\lambda_{ed}}\right)^{N_r}.
\end{multline}
\proof
Invoking \cite[Eq. (1.112.1)]{Tables}, we have
\begin{align} \label{PA:10}
\left(1+\frac{\lambda_2 N_D}{\lambda_1 N_E}\right)^{-N_r}= \left(\frac{ N_E}{\lambda_{ed} N_D}\right)^{N_r} \left(1+O\left(\frac{1}{\lambda_{ed}}\right)\right)^{N_r}=
\left(\frac{ N_E}{\lambda_{ed} N_D}\right)^{N_r}+O\left(\frac{1}{\lambda_{ed}^{N_r+1}}\right),
\end{align}
and
\begin{align} \label{PA:11}
\frac{\lambda_2}{\lambda_1}\left(1+\frac{\lambda_2}{\lambda_1}x\right)^{-(N_r+1)}=\frac{1}{x^{N_r+1}} \left(\frac{1}{\lambda_{ed}}\right)^{N_r} \left(1+O\left(\frac{1}{\lambda_{ed}}\right)\right)^{N_r+1}=\frac{1}{x^{N_r+1}} \left(\frac{1}{\lambda_{ed}}\right)^{N_r} +O\left(\frac{1}{\lambda_{ed}^{N_r+1}}\right).
\end{align}
Then the desired result can be obtained by following similar procedures as in Appendix \ref{proof:theorem:9}.
\endproof
\end{lemma}

Interestingly, we observe that the MRT/MRC scheme achieves a diversity order of $N_r$, which is same as the TZF/MRC scheme.

\section{Numerical Results}  \label{section:5}
In this section, numerical results are presented to illustrate the performance of the proposed proactive eavesdropping schemes and validate the analytical expressions. Unless otherwise specify, the number of transmit and receive antennas at the legitimate monitor is $N_t=N_r=3$, the noise variances at both D and E are normalized such that $N_D=N_E=1$, the self-interference coefficient is $\rho=0.5$, the average channel gains $\lambda_1$, $\lambda_2$, $\lambda_3$ and $\lambda_4$ are set to be 1, 0.1, 0.1 and 1, respectively.

\begin{figure}[htbp]
\centering
\includegraphics[width=4in]{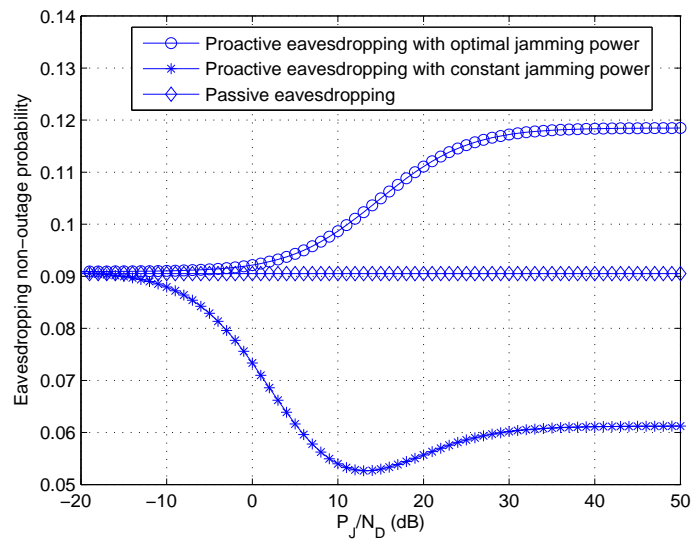}
\caption{Eavesdropping non-outage probability comparison for the SISO case.}
\label{fig:fig2}
\end{figure}

Fig. \ref{fig:fig2} depicts the eavesdropping non-outage probability for the SISO case. For comparison, the performance of the two benchmark schemes proposed in \cite{J.Xu,J.Xu1} are also plotted, namely, 1) Proactive eavesdropping with constant jamming power, i.e., $p_d=P_J$, 2) Passive eavesdropping, i.e., $p_d=0$. As expected, the proposed proactive eavesdropping with optimal jamming power substantially outperforms the other two reference schemes. Moreover, we observe that for the proactive constant-power jamming scheme, increasing the jamming power may decrease the eavesdropping non-outage probability due to the potential severe interference inflicted on the legitimate monitor. In contrast, increasing the maximum jamming power is always beneficial for the proposed proactive eavesdropping scheme with optimal jamming power.

\begin{figure}[htbp]
\centering
\includegraphics[width=4in]{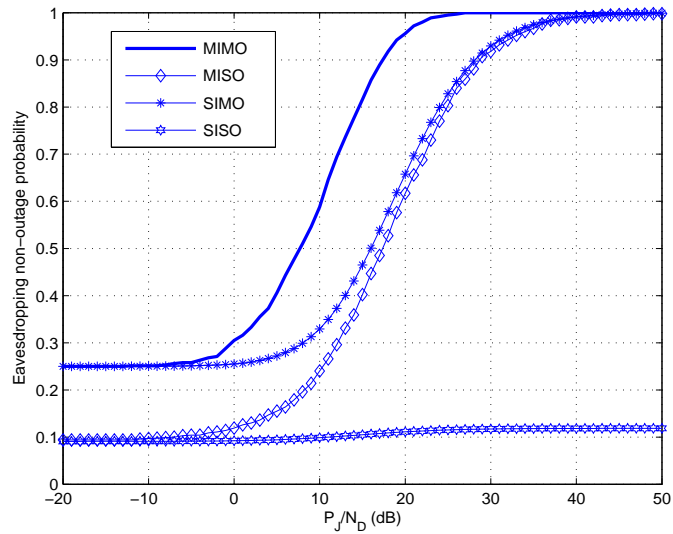}
\caption{Eavesdropping non-outage probability comparison of the MIMO, MISO, SIMO and SISO cases.}
\label{fig:fig3}
\end{figure}

Fig. \ref{fig:fig3} compares the achievable eavesdropping non-outage probability of the MIMO, MISO, SIMO and SISO cases. As expected, the MIMO case always yields the best performance, while the SISO case is the worst. Also, the MISO and SIMO cases significantly outperform the SISO case, thereby demonstrating the potential benefit of implementing multiple antennas at the legitimate monitor. In addition, the performance of SIMO case is in general better then the MISO case. When the maximum jamming power is sufficiently large, the eavesdropping non-outage probability of all multiple antenna cases approaches one. However, if the maximum jamming power is small, the benefit of deploying multiple transmit antenna vanishes.

\begin{figure}[htbp]
\centering
\includegraphics[width=4in]{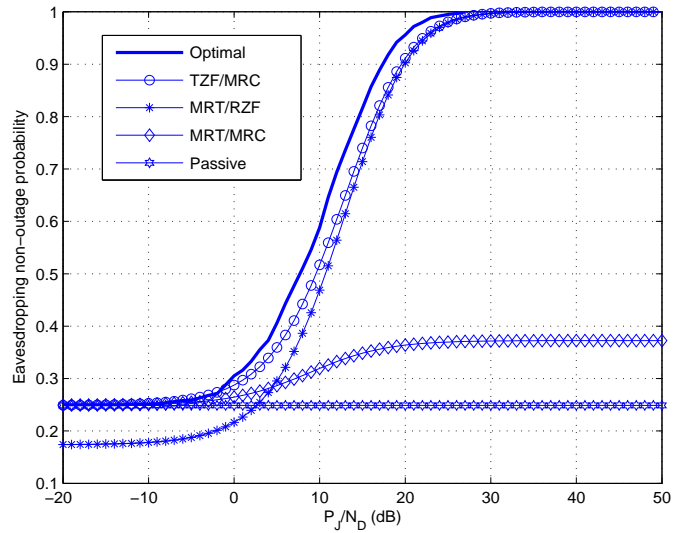}
\caption{Eavesdropping non-outage probability of the MIMO case: Optimal design v.s. Suboptimal design.}
\label{fig:fig4}
\end{figure}

Fig. \ref{fig:fig4} illustrates the eavesdropping non-outage probability of the proposed suboptimal schemes. We observe that, among the proposed suboptimal schemes, the TZF/MRC scheme achieves the best performance, and remarkably, it has a similar performance as the optimal scheme. Also, the performance of the MRT/MRC scheme is noticeably worse than that of the TZF/MRC and MRT/RZF schemes with moderate maximum jamming power, which indicates the critical importance of properly handling the self-interference at the legitimate monitor. In addition, the MRC schemes outperform the RZF scheme at low maximum jamming power region, the reason is that in such region, the self-interference is rather insignificant, hence, it is better to utilize all the receive antennas to enhance the quality of the desired signal, instead of sacrificing one degree of freedom for self-interference suppression.

\begin{figure}[htbp]
\centering
\includegraphics[width=4in]{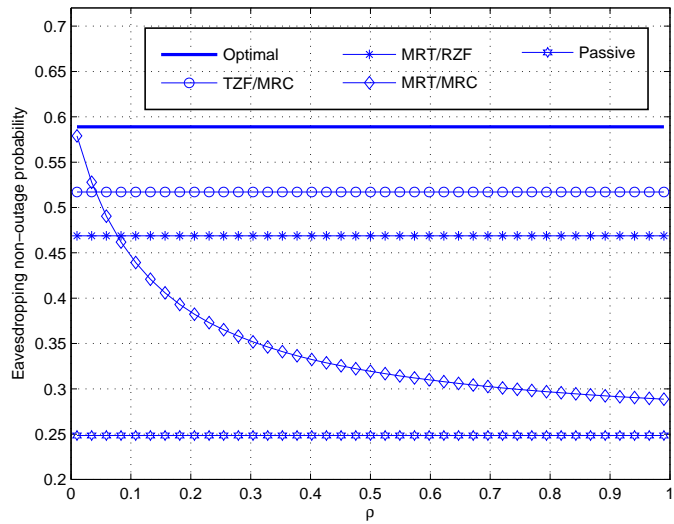}
\caption{Eavesdropping non-outage probability versus self-interference suppression parameter $\rho$ for the MIMO case with $P_J/N_D=10$dB.}
\label{fig:fig5}
\end{figure}

Fig. \ref{fig:fig5} plots the eavesdropping non-outage probability with different self-interference suppression parameter $\rho$ for the MIMO case. We observe that, regardless of $\rho$, the optimal scheme achieves the best performance. Also, for the ZF-based schemes, the eavesdropping non-outage probability remains constant, since both schemes can perfectly eliminate self-interference. While for the MRT/MRC scheme, increasing $\rho$ decreases the eavesdropping non-outage probability, and when $\rho$ is small, the MRT/MRC scheme tends to outperform other suboptimal schemes.

\begin{figure}[htbp]
\begin{minipage}[t]{0.48\linewidth}
\centering
\subfigure[$\lambda_4 = 1$] { \label{fig:fig6a}
\includegraphics[width=3.3in]{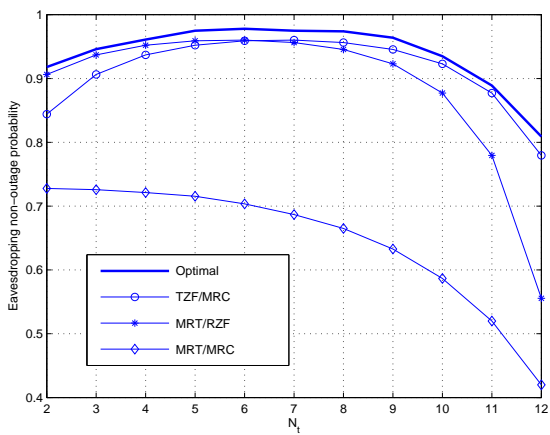}}
\end{minipage}
\begin{minipage}[t]{0.48\linewidth}
\centering
\subfigure[$\lambda_4 = 0.1$] { \label{fig:fig6b}
\includegraphics[width=3.3in]{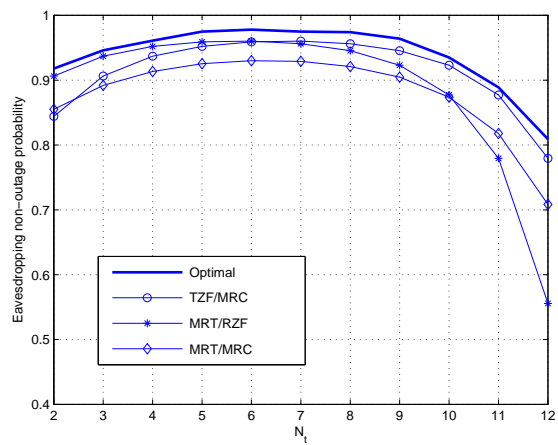}}
\end{minipage}
\caption{Eavesdropping non-outage probability versus $N_t$ for the MIMO case with $N_t+N_r=14$ and $P_J/N_D=10$dB.}
\label{fig:fig6}
\end{figure}

Fig. \ref{fig:fig6} investigates the eavesdropping non-outage probability with different $N_t$ for the MIMO case when $N_t+N_r=14$. From Fig. \ref{fig:fig6a} and Fig. \ref{fig:fig6b}, we see that, for the optimal, TZF/MRC and MRT/RZF schemes, there exists a unique $N_t$ which yields the best performance. However, for the MRT/MRC scheme, the impact of $N_t$ on the achievable performance depends heavily on $\lambda_4$. With large $\lambda_4$, i.e., $\lambda_4=1$, which corresponds to the strong self-interference scenario, it is better to deploy more antennas at the receive side as shown in Fig. \ref{fig:fig6a}. On the other hand, with small $\lambda_4$, i.e., $\lambda_4 = 0.1$, which corresponds to the weak self-interference scenario, the number of transmit and receive antenna needs to be balanced. The main reason is that, with strong self-interference, the benefit of deploying more antennas at the receive side to enhance the eavesdropping channel capacity overweights the capacity degradation of the suspicious channel by employing the same number of transmit antennas.

\begin{figure}[htbp]
\centering
\includegraphics[width=4in]{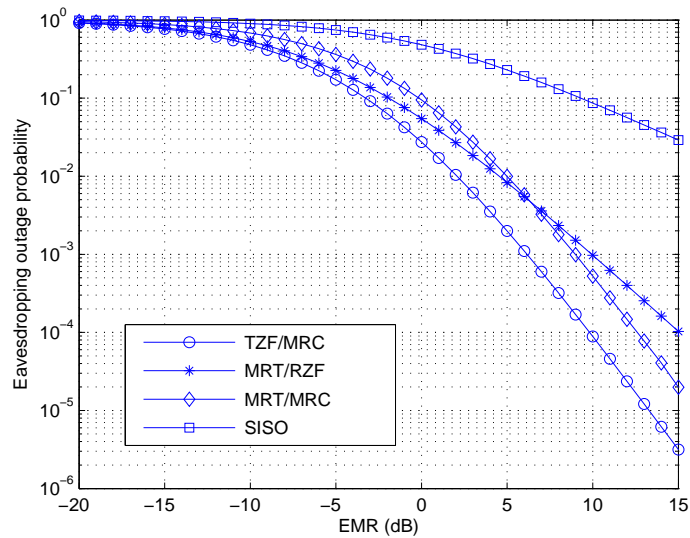}
\caption{Eavesdropping outage probability versus EMR for suboptimal schemes with $P_J/N_D=10$dB.}
\label{fig:fig7}
\end{figure}

Fig. \ref{fig:fig7} examines the eavesdropping outage probability with different EMR for the proposed suboptimal schemes. We observe that both the TZF/MRC and MRT/MRC schemes achieve a diversity order of $N_r$, and the MRT/RZF scheme attains a diversity order of $N_r-1$, while the SISO scheme only achieves unit diversity order, which is consistent with the analytical results presented in section \ref{section:4}. In addition, the MRT/RZF scheme outperforms the MRT/MRC scheme when the EMR is small, while becomes inferior as the EMR increases.

\section{Conclusion}   \label{section:6}
We have studied the joint design of jamming power and transmit/receive beamforming vectors at the legitimate monitor to maximize the eavesdropping non-outage probability. Four different scenarios have been considered. For each scenario, the optimal jamming power was characterized in closed-form. Also, efficient algorithms were proposed to obtain the optimal transmit/receive beamforming vectors. Finally, low-complexity suboptimal beamforming schemes were proposed, and analytical expressions were derived for the achievable eavesdropping non-outage probabilities of the suboptimal schemes. The findings suggest that adopting multiple-antenna tremendously improves the performance of the system. Moreover, the suboptimal TZF/MRC scheme attains similar performance as the optimal scheme, hence provides an attractive low-complexity solution for practical implementation.

\appendices

\section{Proof of Theorem \ref{theorem:3}}\label{proof:theorem:3}
The Lagrangian multiplier function for problem (P9) can be expressed as
\begin{multline} \label{A8:1}
L\left(\mathbf{Z},s,\xi_1,\xi_2,\xi_3,\mathbf{Y}\right)= \rho P_J \text{tr}\left(\mathbf{Z} \mathbf{h}_{ee}^\dagger \mathbf{h}_{ee} \right)+ s N_E - \xi_1 s+\xi_2\left(s-\text{tr} \left(\mathbf{Z}\right)\right)+ \\ \xi_3\left(1-s N_D- P_J \text{tr}
\left(\mathbf{Z} \mathbf{h}_{ed}^\dagger \mathbf{h}_{ed}\right)\right)-\text{tr}\left(\mathbf{Y}\mathbf{Z}\right),
\end{multline}
where $\xi_1 \geq 0$, $\xi_2$ and $\xi_3$ denote the dual variables of problem (P9) associated with the constraints in (19a) and (19c), respectively, while $\mathbf{Y} \succeq \mathbf{0}$ is the matrix dual variable associated with the constraint $\mathbf{Z} \succeq \mathbf{0}$. Since problem (P9) is convex and Slater condition holds true, the Karush-Kuhn-Tucker (KKT) conditions are necessary and sufficient for establishing the optimality \cite{S.Boyd}. The KKT conditions for problem (P9) are
\begin{align}  \label{A8:2}
& \text{constraints (19a)--(19d)} \\
&\xi_1 s=0 \rightarrow \xi_1=0 \tag{48a} \\
&\text{tr}\left(\mathbf{Y}\mathbf{Z}\right)=0 \rightarrow \mathbf{Y}\mathbf{Z}=\mathbf{0} \tag{48b} \\
& \frac{\partial L}{\partial s}= N_E-\xi_1+\xi_2-\xi_3 N_D=0 \tag{48c} \\
& \frac{\partial L}{\partial \mathbf{Z}} = \rho P_J \mathbf{h}_{ee}^\dagger  \mathbf{h}_{ee} -\xi_2 \mathbf{I}_{N_t}-\xi_3  P_J \mathbf{h}_{ed}^\dagger \mathbf{h}_{ed}  -\mathbf{Y} = \mathbf{0}. \tag{48d}
\end{align}
Now, multiplying (48d) by $\mathbf{Z}$ and utilizing (\ref{A8:2}), we have
\begin{align} \label{A8:3}
\rho P_J \text{tr}\left(\mathbf{Z} \mathbf{h}_{ee}^\dagger  \mathbf{h}_{ee} \right) +s N_E=\xi_3,
\end{align}
which implies that $\xi_3 > 0$. Thus, we have
\begin{align}  \label{A8:4}
\mathbf{Y}=\rho P_J \mathbf{h}_{ee}^\dagger  \mathbf{h}_{ee}-\left(\xi_3 N_D-N_E\right)\mathbf{I}_{N_t}-\xi_3 P_J \mathbf{h}_{ed}^\dagger \mathbf{h}_{ed} .
\end{align}
To this end, we find it convenient to give a separate treatment for three different cases depending on the relationship between $\xi_3$, $N_D$ and $N_E$.
\begin{itemize}
\item Case 1: $\xi_3 N_D > N_E$. In this case, $\mathbf{Y}$ is no longer semidefinite positive, which contradicts with the assumption that $\mathbf{Y} \succeq \mathbf{0}$.
\item Case 2: $\xi_3 N_D = N_E$. Noticing that $\mathbf{Y}$ in (\ref{A8:4}) is the difference of two rank-one matrices, it can be readily shown from Weyl's inequalities that $\mathbf{Y}$ cannot be positive-semidefinite except in the case with $\xi_3 P \mathbf{h}_{ed}^\dagger \mathbf{h}_{ed}=0$, i.e., $\xi_3=0$ for non-zero $\mathbf{h}_{ed}$, which contradicts with the assumption $\xi_3 N_D = N_E$.
\item Case 3: $\xi_3 N_D < N_E$. In this case, $\rho P_J\mathbf{h}_{ee}^\dagger  \mathbf{h}_{ee}-\left(\xi_3 N_D-N_E\right)\mathbf{I}_{N_t}$ is a full-rank positive matrix. Multiplying (48d) by $\mathbf{Z}$, we have
    \begin{align}  \label{A8:5}
    \text{rank} \left(\mathbf{Z}\right) &= \text{rank} \left(\left(\rho P_J \mathbf{h}_{ee}^\dagger  \mathbf{h}_{ee}-\left(\xi_3 N_D-N_E\right)\mathbf{I}_{N_t}\right)\mathbf{Z}\right) \nonumber \\
    &= \text{rank}\left(\left(\xi_3 P_J \mathbf{h}_{ed}^\dagger \mathbf{h}_{ed}\right)\mathbf{Z}\right) \leq \min \left(\text{rank}\left(\mathbf{h}_{ed}^\dagger \mathbf{h}_{ed}\right),\text{rank}\left(\mathbf{Z}\right)\right)=1,
    \end{align}
    where the last equality follows from the fact that $\text{rank}\left(\mathbf{Z}\right) \geq 1$ with $\text{tr}\left(\mathbf{Z}\right)=s$. As such, we have $\text{rank}\left(\mathbf{Z}\right) \geq 1$ and $\text{rank}\left(\mathbf{Z}\right) \leq 1$, which implies that $\text{rank}\left(\mathbf{Z}\right)=1$.
\end{itemize}

\section{Proof of Theorem \ref{theorem:4}}\label{proof:theorem:4}
The Lagrangian multiplier function for problem (P15) can be expressed as
\begin{multline} \label{A2:1}
L\left(\mathbf{Z},s,\xi_1,\xi_2,\xi_3,\xi_4,\mathbf{Y}\right)=\frac{\rho P_J}{N_E} \text{tr}\left(\mathbf{Z} \mathbf{H}_{ee}^\dagger \mathbf{h}_{se} \mathbf{h}_{se}^\dagger \mathbf{H}_{ee} \right) - \xi_1 s+\xi_2\left(s-\text{tr} \left(\mathbf{Z}\right)\right)+ \\ \xi_3\left(1-s-\frac{\rho P_J}{N_E} \text{tr}
\left(\mathbf{Z} \mathbf{H}_{ee}^\dagger \mathbf{H}_{ee}\right)\right)+\xi_4\left(s(y-1)-\frac{P_J}{N_D}\text{tr} \left(\mathbf{Z} \mathbf{h}_{ed}^\dagger \mathbf{h}_{ed}\right)\right)-\text{tr}\left(\mathbf{Y}\mathbf{Z}\right),
\end{multline}
where $\xi_1 \geq 0$, $\xi_2$, $\xi_3$ and $\xi_4$ denote the dual variables of problem (P15) associated with the constraints from (27a) to (27d), respectively, while $\mathbf{Y} \succeq \mathbf{0}$ is the matrix dual variable associated with the constraint $\mathbf{Z} \succeq \mathbf{0}$. Since the problem (P15) is convex and Slater condition holds true, the KKT conditions are necessary and sufficient for optimality. The KKT conditions for problem (P15) are
\begin{align}  \label{A2:2}
& \text{constraints (27a)--(27e)} \\
&\xi_1 s=0 \rightarrow \xi_1=0 \tag{53a} \\
&\text{tr}\left(\mathbf{Y}\mathbf{Z}\right)=0 \rightarrow \mathbf{Y}\mathbf{Z}=\mathbf{0} \tag{53b} \\
& \frac{\partial L}{\partial s}= -\xi_1+\xi_2-\xi_3+\xi_4(y-1)=0 \tag{53c} \\
& \frac{\partial L}{\partial \mathbf{Z}} = \frac{\rho P_J}{N_E}\mathbf{H}_{ee}^\dagger \mathbf{h}_{se} \mathbf{h}_{se}^\dagger \mathbf{H}_{ee} -\xi_2 \mathbf{I}_{N_t}-\xi_3 \frac{\rho P_J}{N_E} \mathbf{H}_{ee}^\dagger \mathbf{H}_{ee} -\xi_4 \frac{P_J}{N_D} \mathbf{h}_{ed}^\dagger \mathbf{h}_{ed} -\mathbf{Y} = \mathbf{0}. \tag{53d}
\end{align}
Multiplying (53d) by $\mathbf{Z}$ and utilizing (\ref{A2:2}) yield
\begin{align} \label{A2:3}
\frac{\rho P_J}{N_E} \text{tr}\left(\mathbf{Z} \mathbf{H}_{ee}^\dagger \mathbf{h}_{se} \mathbf{h}_{se}^\dagger \mathbf{H}_{ee} \right)=\xi_3,
\end{align}
which implies $\xi_3 \geq 0$. Thus, we have
\begin{align}  \label{A2:4}
\mathbf{Y}=\frac{\rho P_J}{N_E}\mathbf{H}_{ee}^\dagger \mathbf{h}_{se} \mathbf{h}_{se}^\dagger \mathbf{H}_{ee}+\left(\xi_4(y-1)-\xi_3\right)\mathbf{I}_{N_t}-\xi_3 \frac{\rho P_J}{N_E} \mathbf{H}_{ee}^\dagger \mathbf{H}_{ee} -\xi_4 \frac{P_J}{N_D} \mathbf{h}_{ed}^\dagger \mathbf{h}_{ed}.
\end{align}
We now give a separate treatment for three different cases depending on $\xi_3$ and $\xi_4$.
\begin{itemize}
\item Case 1: $\xi_3 >0$ and $\xi_4 >0$. If $\xi_3 > 0$, exploiting the fact that $\mathbf{Y} \succeq \mathbf{0}$, it is easy to prove that $\xi_4 >0$. If $\frac{\rho P_J}{N_E}\mathbf{H}_{ee}^\dagger \mathbf{h}_{se} \mathbf{h}_{se}^\dagger \mathbf{H}_{ee}+\left(\xi_4(y-1)-\xi_3\right)\mathbf{I}_{N_t}-\xi_3 \frac{\rho P_J}{N_E} \mathbf{H}_{ee}^\dagger \mathbf{H}_{ee}$ is positive definite, multiplying (53d) by $\mathbf{Z}$ yields
    \begin{align}  \label{A2:5}
    \text{rank} \left(\mathbf{Z}\right) &= \text{rank} \left(\left(\frac{\rho P_J}{N_E}\mathbf{H}_{ee}^\dagger \mathbf{h}_{se} \mathbf{h}_{se}^\dagger \mathbf{H}_{ee}+\left(\xi_4(y-1)-\xi_3\right)\mathbf{I}_{N_t}-\xi_3 \frac{\rho P_J}{N_E} \mathbf{H}_{ee}^\dagger \mathbf{H}_{ee}\right)\mathbf{Z}\right) \nonumber \\
    &= \text{rank}\left(\left(\xi_4 \frac{P_J}{N_D} \mathbf{h}_{ed}^\dagger \mathbf{h}_{ed}\right)\mathbf{Z}\right) \leq \min \left(\text{rank}\left(\mathbf{h}_{ed}^\dagger \mathbf{h}_{ed}\right),\text{rank}\left(\mathbf{Z}\right)\right)=1.
    \end{align}

    As for the case where $\frac{\rho P_J}{N_E}\mathbf{H}_{ee}^\dagger \mathbf{h}_{se} \mathbf{h}_{se}^\dagger \mathbf{H}_{ee}+\left(\xi_4(y-1)-\xi_3\right)\mathbf{I}_{N_t}-\xi_3 \frac{\rho P_J}{N_E} \mathbf{H}_{ee}^\dagger \mathbf{H}_{ee}$ is positive semi-definite and its smallest eigenvalue is 0, the optimal rank one solution has been presented in \cite[Appendix B]{G.Zheng}.

\item Case 2: $\xi_3 =0$ and $\xi_4 >0$. In this case, $\frac{\rho P_J}{N_E}\mathbf{H}_{ee}^\dagger \mathbf{h}_{se} \mathbf{h}_{se}^\dagger \mathbf{H}_{ee}+\xi_4(y-1)\mathbf{I}_{N_t}$ is a full-rank positive matrix. Multiplying (53d) by $\mathbf{Z}$, we have
    \begin{align}  \label{A2:6}
    \text{rank} \left(\mathbf{Z}\right) &= \text{rank} \left(\left(\frac{\rho P_J}{N_E}\mathbf{H}_{ee}^\dagger \mathbf{h}_{se} \mathbf{h}_{se}^\dagger \mathbf{H}_{ee}+\xi_4(y-1)\mathbf{I}_{N_t}\right)\mathbf{Z}\right) \nonumber \\
    &= \text{rank}\left(\left(\xi_4 \frac{P_J}{N_D} \mathbf{h}_{ed}^\dagger \mathbf{h}_{ed}\right)\mathbf{Z}\right) \leq \min \left(\text{rank}\left(\mathbf{h}_{ed}^\dagger \mathbf{h}_{ed}\right),\text{rank}\left(\mathbf{Z}\right)\right)=1.
    \end{align}
\item Case 3: $\xi_3 =0$ and $\xi_4 =0$. In this case, $\mathbf{Y}$ reduces to
    \begin{align}  \label{A2:7}
    \mathbf{Y}=\frac{\rho P_J}{N_E}\mathbf{H}_{ee}^\dagger \mathbf{h}_{se} \mathbf{h}_{se}^\dagger \mathbf{H}_{ee},
    \end{align}
    which is a rank-one matrix. Hence, the rank of the optimum $\mathbf{Z}$ is not guaranteed to be one. However, it can be shown that an optimum rank-one matrix can always be recovered from optimum $\mathbf{Z}$ as follows: Suppose the optimum $\mathbf{Z}$ is of rank $r$, i.e., $\mathbf{Z}= \sum_{q=1}^r \sigma_q \mathbf{u}_q \mathbf{u}_q^\dagger$, where $\sigma_q$ and $\mathbf{u}_q$ are the eigenvalues and eigenvectors of the matrix $\mathbf{Z}$, respectively. Due to the equality constraint (27b), $\sum_{q=1}^r \sigma_q=s$. Substituting eigenvalue decomposition of $\mathbf{Z}$ into (53b) yields $\sum_{q=1}^r \sigma_q \mathbf{u}_q^\dagger \mathbf{Y} \mathbf{u}_q =0 $. Recalling that $\mathbf{Y} \succeq \mathbf{0}$, we have
    \begin{align}  \label{A2:8}
   \mathbf{u}_q^\dagger \mathbf{H}_{ee}^\dagger \mathbf{h}_{se} \mathbf{h}_{se}^\dagger \mathbf{H}_{ee} \mathbf{u}_q=0, \forall q
    \end{align}
   From (27c) and (27d), we have
    \begin{align}  \label{A2:9}
    s+ \frac{\rho P_J}{N_E }\sum_{q=1}^r \sigma_q \mathbf{u}_q^\dagger \mathbf{H}_{ee}^\dagger \mathbf{H}_{ee} \mathbf{u}_q =1,
    \end{align}
    and
    \begin{align}  \label{A2:10}
    s(y-1)=\frac{P_J}{N_D} \sum_{q=1}^r \sigma_q \mathbf{u}_q^\dagger \mathbf{h}_{ed}^\dagger \mathbf{h}_{ed} \mathbf{u}_q.
    \end{align}
    Noticing that the objective function of problem (P15) satisfies $\frac{\rho P_J}{N_E} \text{tr}\left(\mathbf{Z} \mathbf{H}_{ee}^\dagger \mathbf{h}_{se} \mathbf{h}_{se}^\dagger \mathbf{H}_{ee} \right)=\xi_3=0$, i.e., $f(y)=0$ in the one dimension search stage, based on (27d), the objective function of problem (P14) becomes $f(y)+\frac{\frac{N_E}{N_D}|h_{sd}|^2}{y}=\frac{\frac{N_E}{N_D}|h_{sd}|^2}{1+\frac{P_J}{sN_D} \text{tr}\left(\mathbf{Z} \mathbf{h}_{ed}^\dagger \mathbf{h}_{ed}\right) }$. To this end, we choose eigenvector such that $\hat{q}=\max_q \mathbf{u}_q^\dagger \mathbf{h}_{ed}^\dagger \mathbf{h}_{ed} \mathbf{u}_q$ and set $\sigma_{\hat{q}}=s$. Due to (\ref{A2:8}), such a choice does not affect the value of $f(y)$, and it is easy to verify that $\sigma_{\hat{q}} \mathbf{u}_{\hat{q}}^\dagger \mathbf{h}_{ed}^\dagger \mathbf{h}_{ed} \mathbf{u}_{\hat{q}} \geq \sum_{q=1}^r \sigma_q \mathbf{u}_q^\dagger \mathbf{h}_{ed}^\dagger \mathbf{h}_{ed} \mathbf{u}_q$, which implies that such a beamforming vector design guarantees a better solution for problem (P14). Therefore, the optimum rank-one matrix recovered from $\mathbf{Z}$ turns out to be $\sigma_{\hat{q}} \mathbf{u}_{\hat{q}} \mathbf{u}_{\hat{q}}^\dagger$.
\end{itemize}

\section{Proof of Theorem \ref{theorem:5}}\label{proof:theorem:5}
The expectation of random variable $X$ is given by ${\tt E}\{X\}=  {\mathop{\rm Prob}\nolimits} \left( \frac{P_S| {h}_{se}|^2}{\rho p_d|{h}_{ee} |^2+N_E} \geq \frac{P_S|h_{sd}|^2}{p_d|{h}_{ed} |^2+N_D} \right)$, which can be computed via
\begin{align} \label{A3:1}
{\tt E}\{X\} = \underbrace{{\mathop{\rm Prob}\nolimits} \left(b<c\right) \times {\mathop{\rm Prob}\nolimits} \left(a<c | b<c\right)}_{I_1}+ \underbrace{{\mathop{\rm Prob}\nolimits} \left( c<b\right)\times {\mathop{\rm Prob}\nolimits} \left(\left. \frac{a-c}{\rho \gamma_{ee}} \leq \frac{P_J}{N_E} \left(b-a\right)\right|c<b\right)}_{I_2},
\end{align}
where $a=\frac{\gamma_{sd}}{\gamma_{se}}$, $b = \frac{\gamma_{ed}}{\rho \gamma_{ee}}$, $c=\frac{N_D}{N_E}$, $\gamma_{sd}=|h_{sd}|^2$, $\gamma_{se}=|h_{se}|^2$, $\gamma_{ed}=|h_{ed}|^2$ and $\gamma_{ee}=|h_{ee}|^2$. Noticing that $\gamma_{sd}$, $\gamma_{se}$, $\gamma_{ed}$ and $\gamma_{ee}$ follow the exponential distribution with mean $\lambda_1$, $\lambda_2$, $\lambda_3$ and $\lambda_4$, respectively, the cumulative distribution function (cdf) of random variables $a$ and $b$ can be derived as
\begin{align}  \label{A3:2}
F_a(x)=\int_0^\infty \!\! \int_0^{xz} \frac{1}{\lambda_1} e^{-\frac{y}{\lambda_1}}  \frac{1}{\lambda_2} e^{-\frac{z}{\lambda_2}} {\rm dydz}=1-\frac{1}{1+\frac{\lambda_2}{\lambda_1}x},
\end{align}
and
\begin{align} \label{A3:3}
F_b(x)=\int_0^\infty \!\! \int_0^{\rho xz} \frac{1}{\lambda_3} e^{-\frac{y}{\lambda_3}}  \frac{1}{\lambda_4} e^{-\frac{z}{\lambda_4}} {\rm dydz}=1-\frac{1}{1+\frac{\rho \lambda_4}{\lambda_3} x}.
\end{align}
Then $I_2$ can be divided into three parts as follows
\begin{multline}  \label{A3:4}
I_2 = \overbrace{{\mathop{\rm Prob}\nolimits} \left(a<c<b\right) \times 1}^{I_3}+\overbrace{{\mathop{\rm Prob}\nolimits} \left( c<a<b\right)\times {\mathop{\rm Prob}\nolimits} \left(\left. \frac{a-c}{\rho \gamma_{ee}} \leq \frac{P_J}{N_E} \left(b-a\right)\right|c<a<b\right)}^{I_4}+ \\ \underbrace{{\mathop{\rm Prob}\nolimits} \left(c<b<a\right) \times 0}_{I_5}.
\end{multline}
$I_1$ and $I_3$ can be evaluated by
\begin{align} \label{A3:5}
I_1+I_3=F_a(c)F_b(c)+F_a(c)\left(1-F_b(c)\right) =1-\frac{1}{1+\frac{\lambda_2 N_D}{\lambda_1 N_E}}.
\end{align}
The next step is to calculate $I_4$, and we have
\begin{align} \label{A3:6}
I_4={\mathop{\rm Prob}\nolimits} \left(\frac{\frac{\gamma_{sd}}{\gamma_{se}}N_E-N_D}{\gamma_{ed}-\rho \frac{\gamma_{ee} \gamma_{sd}}{\gamma_{se}}} \leq P_J,\frac{N_D}{N_E}<\frac{\gamma_{sd}}{ \gamma_{se}}<\frac{\gamma_{ed}}{\rho \gamma_{ee}}\right).
\end{align}
Averaging over $\gamma_{ed}$ and $\gamma_{ee}$, we obtain
\begin{align} \label{A3:7}
I_4= \int_{\frac{N_D}{N_E}}^\infty e^{-\frac{x N_E-N_D}{\lambda_3 P_J}} \frac{1}{1+\frac{\rho \lambda_4}{\lambda_3} x} f_a(x) \rm dx,
\end{align}
where $f_a(x)$ is the probability density function (pdf) of random variable $a$. Invoking \cite[Eq. (3.353.1)]{Tables} and \cite[Eq. (3.352.2)]{Tables}, $I_4$ can be computed as
\begin{multline} \label{A3:8}
I_4=\frac{\rho \lambda_1\lambda_2\lambda_3\lambda_4}{\left(\rho\lambda_1\lambda_4-\lambda_2\lambda_3\right)^2} \exp \left(\frac{N_D}{\lambda_3 P_J}+\frac{N_E}{\rho \lambda_4 P_J}\right)\Gamma \left(0,\frac{N_D}{\lambda_3 P_J}+\frac{N_E}{\rho \lambda_4 P_J}\right) -\frac{\lambda_1\lambda_2\lambda_3 N_E}{(\rho \lambda_1\lambda_4-\lambda_2\lambda_3)(\lambda_1N_E+\lambda_2N_D)}\\+\left(\frac{\lambda_1N_E}{P_J(\rho\lambda_1\lambda_4-\lambda_2\lambda_3)}-\frac{\rho \lambda_1\lambda_2\lambda_3\lambda_4}{\left(\rho\lambda_1\lambda_4-\lambda_2\lambda_3\right)^2}\right) \exp \left(\frac{N_D}{\lambda_3 P_J}+\frac{\lambda_1N_E}{\lambda_2 \lambda_3 P_J}\right) \Gamma \left(0,\frac{N_D}{\lambda_3 P_J}+\frac{\lambda_1N_E}{\lambda_2 \lambda_3 P_J}\right) .
\end{multline}
To this end, the desired result can be obtained along with some simple algebraic manipulations.

\section{Proof of Lemma \ref{lemma:3}}\label{proof:lemma:3}
Based on (\ref{A3:5}) and (\ref{A3:7}), we have
\begin{align}  \label{A4:1}
P_{\sf out}=\frac{1}{1+\frac{\lambda_2 N_D}{\lambda_1 N_E}}-\int_{\frac{N_D}{N_E}}^\infty e^{-\frac{x N_E-N_D}{\lambda_3 P_J}} \frac{1}{1+\frac{\rho \lambda_4}{\lambda_3} x} \frac{\lambda_1\lambda_2}{\left(\lambda_1+\lambda_2 x\right)^2} \rm dx.
\end{align}
When $\lambda_{ed} \rightarrow \infty$, invoking \cite[Eq. (1.112.1)]{Tables} and \cite[Eq. (1.112.2)]{Tables}, we have
\begin{align} \label{A4:2}
\frac{1}{1+\frac{\lambda_2 N_D}{\lambda_1 N_E}} = \frac{ N_E}{\lambda_{ed} N_D} + O\left(\frac{1}{\lambda_{ed}^2}\right),
\end{align}
and
\begin{align} \label{A4:3}
\frac{\lambda_1\lambda_2}{\left(\lambda_1+\lambda_2 x\right)^2}= \frac{1}{\lambda_{ed} x^2}+O\left(\frac{1}{\lambda_{ed}^2}\right).
\end{align}
As such, in the high SNR regime, $P_{\sf out}$ reduces to
\begin{align} \label{A4:4}
P_{\sf out}^\infty=\frac{N_E}{\lambda_{ed} N_D}-\int_{\frac{N_D}{N_E}}^\infty e^{-\frac{x N_E-N_D}{\lambda_3 P_J}} \frac{1}{1+\frac{\rho \lambda_4}{\lambda_3} x} \frac{1}{\lambda_{ed} x^2} \rm dx.
\end{align}
To this end, utilizing \cite[Eq. (3.353.1)]{Tables} and \cite[Eq. (3.351.4)]{Tables} yields the desired result.

\section{Proof of Theorem \ref{theorem:6}} \label{proof:theorem:6}
Define $ \gamma_{se}=||\mathbf{h}_{se}||^2$ and $\gamma_{ed}=|\mathbf{h}_{ed} \mathbf{\Pi}_1 \mathbf{h}_{ed}^\dagger|$, then it is easy to show that $\gamma_{se}$ follows the chi-square distribution with $2N_r$ degrees of freedom, with pdf given by \cite{M.K.Simon}
\begin{align} \label{A5:1}
f_{se}(x)=\frac{x^{N_r-1}}{\lambda_2^{N_r} \Gamma(N_r)}e^{-\frac{x}{\lambda_2}}.
\end{align}
Also, the pdf of $\gamma_{ed}$ is given by \cite{M.Mohammadi}
\begin{align} \label{A5:2}
f_{ed}(x)=\frac{x^{N_t-2}}{\lambda_3^{N_t-1} \Gamma(N_t-1)}e^{-\frac{x}{\lambda_3}}.
\end{align}
As such, the eavesdropping non-outage probability can be written as
\begin{align} \label{A5:3}
{\tt E}\{X\}={\mathop{\rm Prob}\nolimits} \left( \frac{P_S}{N_E} \gamma_{se} \geq \frac{\frac{P_S}{N_D}\gamma_{sd}}{\frac{P_J}{N_D}\gamma_{ed}+1} \right).
\end{align}
Conditioned on $\gamma_{se}$ and $\gamma_{ed}$, utilizing \cite[Eq. (3.351.1)]{Tables}, we obtain
\begin{align} \label{A5:4}
 {\tt E}\{X\} = 1- \exp \left(-\frac{\gamma_{se}N_D}{\lambda_1 N_E}\left(\frac{P_J}{N_D}\gamma_{ed}+1\right)\right).
\end{align}
Averaging over $\gamma_{se}$, with the help of \cite[Eq. (3.351.3)]{Tables}, we have
\begin{align} \label{A5:5}
 {\tt E}\{X\} = 1- \left(1+\frac{\lambda_2 N_D}{\lambda_1 N_E} \left(\frac{P_J}{N_D}\gamma_{ed}+1\right)\right)^{-N_r}.
\end{align}
Finally, substituting (\ref{A5:2}) into (\ref{A5:5}) yields
\begin{align} \label{A5:6}
  {\tt E}\{X\} = 1- \int_0^\infty \frac{\left(\frac{ \lambda_1 N_E}{\lambda_2 P_J}\right)^{N_r}}{\left(x+\frac{N_D}{P_J}+\frac{\lambda_1 N_E}{\lambda_2 P_J}\right)^{N_r}}  \frac{x^{N_t-2}}{\lambda_3^{N_t-1} \Gamma(N_t-1)}e^{-\frac{x}{\lambda_3}}  \rm dx .
\end{align}
Making a change of variable $t=x+\frac{N_D}{P_J}+\frac{\lambda_1 N_E}{\lambda_2 P_J}$ and applying the binomial expansion, the desired result can be obtained  with the help of \cite[Eq. (3.381.3)]{Tables}.

\section{Proof of Lemma \ref{lemma:4}}\label{proof:lemma:4}
Starting from (\ref{A5:5}), conditioned on $\gamma_{ed}$, utilizing \cite[Eq. (1.112.1)]{Tables} yields
\begin{align} \label{A6:1}
P_{\sf out}=\left(\frac{N_E}{N_D\left(1+\frac{P_J}{N_D}\gamma_{ed}\right)\lambda_{ed}}\right)^{N_r} \left(1+O\left(\frac{1}{\lambda_{ed}}\right)\right)^{N_r}.
\end{align}
Omitting the high order items, we obtain
\begin{align} \label{A6:2}
P_{\sf out}^\infty=\left(\frac{N_E}{N_D\left(1+\frac{P_J}{N_D}\gamma_{ed}\right)\lambda_{ed}}\right)^{N_r}.
\end{align}
To this end, invoking \cite[Eq. (3.381.3)]{Tables}, the desired result can be obtained.

\section{Proof of Theorem \ref{theorem:9}} \label{proof:theorem:9}
Define $\gamma_{ee}=\frac{|\mathbf{h}_{se}^\dagger \mathbf{H}_{ee} \mathbf{h}_{ed}^\dagger|^2}{||\mathbf{h}_{se}||^2||\mathbf{h}_{ed}||^2}$, then the key task is to derive the pdf of $\gamma_{ee}$, which we do in the following. Let $\mathbf{p}=\frac{\mathbf{H}_{ee} \mathbf{h}_{ed}^\dagger}{||\mathbf{h}_{ed}||}$, then according to \cite{A.Shah}, the elements of $N_r \times 1$ vector $\mathbf{p}$ are i.i.d. zero-mean complex Gaussian random variables with variance $\lambda_4$, and $\mathbf{p}$ is independent of $\mathbf{h}_{ed}$. Now consider the scalar $q=\frac{\mathbf{h}_{se}^\dagger \mathbf{p}}{||\mathbf{h}_{se}||}$, we have
\begin{align}  \label{A7:3}
{\tt E}\{q|\mathbf{h}_{se}\}= \frac{\mathbf{h}_{se}^\dagger}{||\mathbf{h}_{se}||} {\tt E}\{\mathbf{p}\}=0,
\end{align}
and
\begin{align}  \label{A7:4}
{\tt E}\{|q|^2|\mathbf{h}_{se}\}= \frac{\mathbf{h}_{se}^\dagger {\tt E}\{ \mathbf{p} \mathbf{p}^\dagger \} \mathbf{h}_{se} }{||\mathbf{h}_{se}||^2}= \lambda_4  \frac{\mathbf{h}_{se}^\dagger \mathbf{I}_{N_r} \mathbf{h}_{se} }{||\mathbf{h}_{se}||^2}=\lambda_4,
\end{align}
which imply that $q$ is a zero-mean complex Gaussian random variable with variance $\lambda_4$. Therefore, $\gamma_{ee}=q^*q$ follows an exponential distribution with mean $\lambda_4$.

Then, we have
\begin{align}  \label{A7:5}
F_a(x)=\int_0^\infty \!\! \int_0^{xz} \frac{1}{\lambda_1} e^{-\frac{y}{\lambda_1}}  \frac{z^{N_r-1}}{\lambda_2^{N_r} \Gamma(N_r)} e^{-\frac{z}{\lambda_2}} {\rm dydz}=1-\frac{1}{\left(1+\frac{\lambda_2}{\lambda_1}x\right)^{N_r}},
\end{align}
and
\begin{align}  \label{A7:6}
F_b(x)=\int_0^\infty \!\! \int_0^{\rho xz} \frac{y^{N_t-1}}{\lambda_3^{N_t} \Gamma(N_t)} e^{-\frac{y}{\lambda_3}}  \frac{1}{\lambda_4} e^{-\frac{z}{\lambda_4}} {\rm dydz}=
1-\sum_{k=0}^{N_t-1} \frac{\left(\frac{\rho \lambda_4}{\lambda_3}x\right)^k}{\left(1+\frac{\rho \lambda_4}{\lambda_3} x\right)^{k+1}}.
\end{align}
Therefore,
\begin{align} \label{A7:7}
I_1+I_3=1-\frac{1}{\left(1+\frac{\lambda_2 N_D}{\lambda_1 N_E}\right)^{N_r}}.
\end{align}
Invoking \cite[Eq. (3.353.2)]{Tables} and \cite[Eq. (3.352.3)]{Tables}, $I_4$ can be computed as
\begin{align}  \label{A7:8}
I_4=\sum_{k=0}^{N_t-1} \sum_{m=0}^k \frac{1}{(k-m)!} \int_{\frac{N_D}{N_E}}^\infty \left(\frac{x N_E-N_D}{\lambda_3P_J}\right)^{k-m} e^{-\frac{x N_E-N_D}{\lambda_3P}}
\frac{\left(\frac{\rho\lambda_4}{\lambda_3}x\right)^m}{\left(1+\frac{\rho \lambda_4}{\lambda_3}x\right)^{m+1}} \frac{N_r \frac{\lambda_2}{\lambda_1}}{\left(1+\frac{\lambda_2}{\lambda_1}x\right)^{N_r+1}} \rm dx.
\end{align}
To this end, pulling everything together yields the desired result.

\begin{footnotesize}

\end{footnotesize}


\begin{thebibliography}{1}

\bibitem{Z.Ding}
Z. Ding, K. K. Leung, D. L. Goeckel, and D. Towsley, ``On the application of cooperative transmission to secrecy communiactions,'' {\em IEEE J. Sel. Areas Commun.}, vol. 30, no. 2, pp. 359--368, Feb. 2012.

\bibitem{N.Yang}
N. Yang, H. A. Suraweera, I. B. Collings, and C. Yuen, ``Physical layer security of TAS/MRC with antenna correlation,'' {\em IEEE Trans. Inf. Foren. Sec.}, vol. 8, no. 1, pp. 254--259, Jan. 2013.

\bibitem{J.Zhu1}
J. Zhu, R. Schober, and V. K. Bhargava, ``Secure transmission in multicell massive MIMO systems,'' {\em IEEE Trans. Wireless Commun.}, vol. 13, no. 9, pp. 4766--4781, Sep. 2014.

\bibitem{F.Zhu}
F. Zhu, F. Gao, M. Yao, and H. Zou, ``Joint information- and jamming-beamforming for physcial layer security with full duplex base station,''
{\em IEEE Trans. Signal Process.}, vol. 62, no. 24, pp. 6391--6401, Dec. 2014.

\bibitem{Y.Zou3}
Y. Zou, J. Zhu, X. Wang, and V. Leung, ``Improving physical-layer security in wireless communications using diversity techniques,'' {\em IEEE Network}, vol. 29, no. 1, pp. 42--48, Jan. 2015.

\bibitem{F.Qahtani}
F. Al-Qahtani, C. Zhong, and H. M. Alnuweiri, ``Opportunistic relay selection for secrecy enhancement in cooperative networks,'' {\em IEEE Trans. Commun.}, vol. 63, no. 5, pp. 1756--1770, May 2015.

\bibitem{X.Fang}
X. Fang, X. Sha, and L. Mei, ``Guaranteeing wireless communication secrecy via a WFRFT-based cooperative system,'' {\em China Commun.}, vol. 12, no. 9, pp. 76--82, Sep. 2015.

\bibitem{J.Zhu2}
J. Zhu, R. Schober, and V. K. Bhargava, ``Linear procoding of data and artificial noise in secure massive MIMO systems,'' {\em IEEE Trans. Wireless Commun.}, vol. 15, no. 3, pp. 2245--2261, Mar. 2016.

\bibitem{S.Gong}
S. Gong, C. Xing, Z. Fei, and J. Kuang, ``Resource allocation for physical layer security in heterogeneous network with hidden eavesdropper,'' {\em China Commun.}, vol. 13, no. 3, pp. 82--95, Mar. 2016.

\bibitem{X.Jiang}
X. Jiang, C. Zhong, X. Chen, T. Q. Duong, T. A. Tsiftsis, and Z. Zhang, ``Secrecy performance of wirelessly powered wiretap channels,'' {\em IEEE Trans. Commun.}, vol. 64, no. 9, pp. 3858--3871, Sep. 2016.

\bibitem{Y.Zou2}
Y. Zou, J. Zhu, X. Wang, and L. Hanzo, ``A survey on wireless security: Technical challenges, recent advances, and future trends,'' {\em Proc. IEEE}, vol. 104, no. 9, pp. 1727--1765, Sep. 2016.

\bibitem{Y.Huang}
Y. Huang, J. Wang, C. Zhong, T. Q. Duong, and G. K. Karagiannidis, ``Secure transmission in cooperative relaying networks with multiple antennas,'' {\em IEEE Trans. Wireless Commun.}, vol. 15, no. 10, pp. 6843--6856, Oct. 2016.

\bibitem{X.Jiang1}
X. Jiang, C. Zhong, Z. Zhang, and G. K. Karagiannidis, ``Power beacon assisted wiretap channels with jamming,'' {\em IEEE Trans. Wireless Commun.}, vol. 15, no. 12, pp. 8353--8367, Dec. 2016.

\bibitem{J.Zhu3}
J. Zhu, W. Xu, and N. Wang, ``Secure massive MIMO systems with limited RF chains,'' accepted to appear in {\em IEEE Trans. Veh. Technol.}, 2016.

\bibitem{J.Zhu4}
J. Zhu, D. Ng, N. Wang, R. Schober, and V. Bhargava, ``Analysis and design of secure massive MIMO systems in the presence of hardware impairments,'' accepted to appear in {\em IEEE Trans. Wireless Commun.}, 2017.

\bibitem{S.Goel}
S. Goel and R. Negi, ``Guaranteeing secrecy using artificial noise,'' {\em IEEE Trans. Wireless Commun.}, vol. 7, no. 6, pp. 2180--2189, June 2008.

\bibitem{X.Zhou}
X. Zhou and M. McKay, ``Secure transmission with artificial noise over fading channels: Achievable rate and optimal power allocation,'' {\em IEEE Trans. Veh. Technol.}, vol. 59, no. 8, pp. 3831--3842, Oct. 2010.

\bibitem{A.Mukherjee}
A. Mukherjee and A. Swindlehurst, ``Robust beamforming for security in MIMO wiretap channels with imperfect CSI,''
{\em IEEE Trans. Signal Process.}, vol. 59, no. 1, pp. 351--361, Jan. 2011.

\bibitem{C.Jeong}
C. Jeong, I. Kim, and D. Kim, ``Joint secure beamforming design at the source and the relay for an amplify-and-forward MIMO untrusted relay system,''
{\em IEEE Trans. Signal Process.}, vol. 60, no. 1, pp. 310--325, Jan. 2012.

\bibitem{J.Xu1}
J. Xu, L. Duan, and R. Zhang, ``Proactive eavesdropping via cognitive jamming in fading channels,'' in {\em Proc. IEEE ICC}, Kuala Lumpur, Malaysia, May 2016, pp. 1--6.

\bibitem{J.Xu}
J. Xu, L. Duan, and R. Zhang, ``Proactive eavesdropping via jamming for rate maximization over Rayleigh fading channels,'' {\em IEEE Wireless Commun. Lett.}, vol. 5, no. 1, pp. 80--83, Feb. 2016.

\bibitem{Y.Zeng}
Y. Zeng and R. Zhang, ``Wireless information surveillance via proactive eavesdropping with spoofing relay,'' {\em IEEE J. Sel. Topics Signal Process.}, vol. 10, no. 8, pp. 1449--1461, Dec. 2016.

\bibitem{G.Zheng}
G. Zheng, I. Krikidis, J. Li, A. P. Petropulu, and B. Ottersten, ``Improving physical layer security using full-duplex jamming receivers,''
{\em IEEE Trans. Signal Process.}, vol. 61, no. 20, pp. 4962--4974, Oct. 2013.

\bibitem{Tables}
I. S. Gradshteyn and I. M. Ryzhik, {\em Table of Integrals, Series and Products,} $6$th Ed.. San Diago: Academic Press, 2000.

\bibitem{T.Riihonen}
T. Riihonen, S. Werner, and R. Wichman, ``Mitigation of loopback self-interference in full-duplex MIMO relays,''
{\em IEEE Trans. Signal Process.}, vol. 59, no. 12, pp. 5983--5993, Dec. 2011.

\bibitem{R.A.Horn}
R. A. Horn and C. A. Johnson, {\em Matrix Analysis.} Cambridge Univ. Press, 2nd Ed., New York, NY: 2013.

\bibitem{W.W.Hager}
W. W. Hager, ``Updating the inverse of a matrix,'' {\em SIAM Review,} vol. 31, no. 2, pp. 221--239, June 1989.

\bibitem{A.Charnes}
A. Charnes and W. W. Cooper, ``Programming with linear fractional functionals,'' {\em Naval Res. Logist. Quarter.}, vol. 9, pp. 181-186, Dec. 1962.

\bibitem{S.Boyd}
S. Boyd and L. Vandenberghe, {\em Convex Optimization.} Cambridge, U.K.: Cambridge Univ. Press, 2004.

\bibitem{Z.Zhang}
Z. Zhang, K. C. Teh, and K. H. Li, ``A semidefinite relaxation approach for beamforming in cooperative clustered multicell systems with novel limited feedback scheme,'' {\em IEEE Trans. Veh. Technol.}, vol. 63, no. 4, pp. 1740--1748, May 2014.

\bibitem{Q.Li}
Q. Li and W. Ma, ``Optimal and robust transmit designs for MISO channel secrecy by semidefinite programming,''
{\em IEEE Trans. Signal Process.}, vol. 59, no. 8, pp. 3799--3812, Aug. 2011.

\bibitem{J.Zhou}
J. Zhou, R. Cao, H. Gao, C. Zhang, and T. Lv, ``Secure beamforming and artificial noise design in interference networks with imperfect ECSI,'' in Proc. {\em IEEE ICCW Phys. Layer Secur.}, pp. 423--428, London, UK, June 2015.

\bibitem{Y.Zou}
Y. Zou, X. Wang, and W. Shen, ``Optimal relay selection for physical-layer security in cooperative wireless networks,'' {\em IEEE J. Sel. Areas Commun.}, vol. 31, no. 10, pp. 2099--2111, Oct. 2013.

\bibitem{Y.Zou1}
Y. Zou, X. Li, and Y. Liang, ``Secrecy outage and diversity analysis of cognitive radio systems,'' {\em IEEE J. Sel. Areas Commun.}, vol. 32, no. 11, pp. 2222--2236, Nov. 2014.

\bibitem{L.Fan}
L. Fan, S. Zhang, T. Q. Duong, and G. K. Karagiannidis, ``Secure switch-and-stay combining (SSSC) for cognitive relay networks,'' {\em IEEE Trans. Commun.}, vol. 64, no. 1, pp. 70--82, Jan. 2016.

\bibitem{J.Zhu}
J. Zhu, Y. Zou, G. Wang, Y. Yao, and G. K. Karagiannidis, ``On secrecy performance of antenna-selection-aided MIMO systems against eavesdropping,'' {\em IEEE Trans. Veh. Technol.}, vol. 65, no. 1, pp. 214--225, Jan. 2016.

\bibitem{H.A.Suraweera}
H. A. Suraweera, I. Krikidis, G. Zheng, C. Yuen, and P. J. Smith, ``Low-complexity end-to-end performance optimization in MIMO full-duplex relay systems,'' {\em IEEE Trans. Wireless Commun.}, vol. 13, no. 2, pp. 913--927, Feb. 2014.

\bibitem{M.Mohammadi}
M. Mohammadi, B. K. Chalise, H. A. Suraweera, C. Zhong, G. Zheng, and I. Krikidis, ``Throughput analysis and optimization of wireless-powered multiple antenna full-duplex relay systems,'' {\em IEEE Trans. Commun.}, vol. 64, no. 4, pp. 1769--1785, Apr. 2016.

\bibitem{M.K.Simon}
M. K. Simon and M. S. Alouini, ``Digital Communication over Fading Channels: A Unified Approach to Performance Analysis,'' Hoboken, NJ: Wiley, 2000.

\bibitem{A.Shah}
A. Shah and A. M. Haimovich, ``Performance analysis of maximal ratio combining and comparison with optimum combining for mobile radio communications with cochannel interference,'' {\em IEEE Trans. Veh. Technol.}, vol. 49, no. 4, pp. 1454--1463, July 2000.

\end{thebibliography}
\end{document}